\documentclass{sig-alternate-05-2015}

\usepackage{subfigure}
\usepackage{graphicx}
\usepackage{bbm}
\begin{document}


\title{Spatial {\ttlit multi-LRU} Caching for Wireless Networks with Coverage Overlaps}
\numberofauthors{2}
\author{
\alignauthor 
Anastasios Giovanidis\\
	\affaddr{CNRS/T\'el\'ecom ParisTech - LTCI Lab}\\
	\affaddr{23, av. d'Italie, 75013, Paris, France}\\
	\email{giovanid@enst.fr}
\alignauthor 
Apostolos Avranas\\
	\affaddr{T\'el\'ecom ParisTech - LTCI Lab, France \&}\\ 
	\affaddr{AUTh, Thessaloniki, Greece}\\
	\email{eavranasa@gmail.com}
}
%
\date{10 novembre 2015}

\maketitle

\begin{abstract}
This article introduces a novel family of decentralised caching policies, applicable to wireless networks with finite storage at the edge-nodes (stations). These policies are based on the \textit{Least-Recently-Used} replacement principle, and are, here, referred to as spatial \textit{multi-LRU}. Based on these, cache inventories are updated in a way that provides content diversity to users who are covered by, and thus have access to, more than one station. Two variations are proposed, namely the \textit{multi-LRU-One} and \textit{-All}, which differ in the number of replicas inserted in the involved caches. By introducing spatial approximations, we propose a Che-like method to predict the hit probability, which gives very accurate results under the Independent Reference Model (IRM). It is shown that the performance of multi-LRU increases the more the multi-coverage areas increase, and it approaches the performance of other proposed centralised policies, when multi-coverage is sufficient. For IRM traffic multi-LRU-One outperforms multi-LRU-All, whereas when the traffic exhibits temporal locality the -All variation can perform better. 
\end{abstract}


\category{C.2.1}{Network Architecture and Design}[Wireless communication, Network topology, Distributed networks]
\terms{Performance}


\section{Introduction}

The design of today's and future networks is characterised by a paradigm shift, from a host-centric communication architecture, towards an Information Centric Networking (ICN) one. The focus is on information itself, and how this can be best accessed \cite{ICNkatsaros14}. Within this setting, network nodes are equipped with storage capacity, where data objects can be temporarily cached. In this way, information can be made available close to the user, it can be retrieved with minimum delay, and possibly with a quality adaptable to the users' preferences, as envisioned for example in cases of multimedia files. The principal benefit of the approach is the reduction of traffic flow at the core network, by serving demands from intermediate nodes \cite{Sourl11}. This further results in congestion avoidance and a better exploitation of the backbone resources.

The edge-nodes constitute a very important part of the architecture, since it is where the users directly have access to. When these nodes are equipped with storage capability, so that users can retrieve their data objects directly from them, average download path length can be minimised \cite{FayaSIG13}. Caching at the edge definitely offers a potential increase in performance of ICNs, it comes however at the cost of a distributed implementation and management over a very vast area, where edge nodes are placed. If these nodes are chosen to be the base stations and small cells of a heterogeneous network \cite{GolrezaeiINFOCOM12, BastugEURASIP15}, it is fairly clear that thousands of nodes within each city are considered, a number which increases by a factor of hundred (or more) if user equipment and other devices are also included as having storage potential. The large number of nodes, together with the relatively small memory size installed on each one, creates big challenges related to their cache management.

We consider the wireless edge of a content centric network, which consists of a set of transmitting nodes taking fixed positions on a planar area, and a set of users dynamically arriving at this area and asking for service. The set of transmitters can refer to base stations (BSs) of a cellular network, small stations of heterogeneous networks, WIFI hotspots, or any other type of wireless nodes that can provide access to an arriving user who demands for a specific data object. A user can be covered by multiple of these nodes, but he/she will \textit{choose only one} to be served from. All nodes are equipped with memory of size $K$ objects, which offers the possibility to cache a fraction of the existing data. When the user's request is found in the cache of some covering station, then the user is served directly by this one. Otherwise, the request is retrieved from the core network. 

An important question is \textit{how to maximise the hit probability}, by managing the available edge-memories. The \textit{hit probability}, is defined as the probability that a user will find her/his demand cached in the memory of one of the cells she/he is covered from. By \textit{managing}, we mean to decide on: Which objects to install in each cache? How to update the cache inventories over time?  


Given the possibility for multi-coverage, cache management should target two, not necessarily conflicting, goals: On the one hand make popular objects, requested by the large bulk of demands, generously available at many geographical locations. On the other, make good use of multi-coverage, by filling the memory caches in a way that a user has access to as many different objects as possible, so that also less popular contents are served directly by the caches. Additionally, since - as explained above - wireless nodes (BSs) are scattered over a very large area and are of considerable number, related operations should be distributed as in \cite{MassoulieCache15, BorstINFOCOM10, LeconteITC15}, and centralised solutions should be avoided.


\subsection{Related Research} There exists a variety of cache placement policies that apply to \textit{single caches}, when no coverage overlap is considered. These include the Least Frequently Used (LFU), the Least Recently Used (LRU), and their variations. Specifically LRU has been extensively studied and approximations to the hit probability have been proposed, like the one from Dan and Towsley \cite{DanTow90}. Che et al proposed in 2002 \cite{CheApprox02} a decomposition and a simple approximation for the single-LRU under the Independent Reference Model (IRM) \cite{CoffmanBook73}, which results in an analytical formula for the hit probability with excellent fit to simulations. This fitness is theoretically explained by Fricker et al in \cite{FriRoRo12}. Application of the Che approximation under more general traffic conditions, to variations of the LRU for single caches as well as networks of caches, is provided by Martina et al \cite{LeoINFOCOM14}. In that work, and further in Elayoubi and Roberts \cite{ElRobSIGCOMM15}, it is shown that for mobile networks, application of pre-filtering improves the performance of LRU. 

There can be strong dependencies between content demands, objects can have a finite lifespan, and new ones can appear anytime. These phenomena constitute the \textit{temporal locality}, not captured from the IRM model. Such type of traffic was studied for LRU initially by Jelenkovi{\'c} and Radovanovi{\'c} \cite{JelenDep04}, and recently using also statistics from user measurements, by Traverso et al \cite{TraversoTranMult15} and Olmos et al \cite{OlmosTEMPO14}.

The problem of optimal content placement, when network areas are covered by more than one station has also been recently studied in the literature. A number of pro-active caching policies have been proposed, where the cache inventories are pre-filled by content, based on knowledge of the content popularity distribution and additional network-related information. Golrezaei et al \cite{GolrezaeiINFOCOM12} find the optimal content placement that maximises hit probability, when full network information (popularity, node and user positions) is available. They formulate a binary optimisation problem and propose approximation and greedy algorithms for its solution. Using reduced information (content popularity, coverage probability), B{\l}aszczyszyn and Giovanidis \cite{BlaGioICC15} provide a randomised strategy that maximises the hit probability. Poularakis et al. \cite{PoulTCOM14} formulate and solve the joint content placement and user association problem that maximises the fraction of content served by the caches of the edge-nodes. Araldo et al. \cite{AraldoSIGC14} propose joint cache sizing/object placement/path selection policies that consider also the cost of content retrieval. Recently, Naveen et al. \cite{MassoulieCache15} have formulated the problem in a way to include the bandwidth costs, and have proposed an online algorithm for its solution. Further distributed replication strategies that use different system information are proposed by Borst et al \cite{BorstINFOCOM10}, and also by Leconte et al \cite{LeconteITC15}. The problem of optimal request routing and content caching for minimum average content access delay in heterogeneous networks is studied by Dehghan et al in \cite{DehgINFO15}.

The cache management problem for cellular networks has also been approached using point process modelling of the network node positions. Bastug et al. \cite{BastugEURASIP15} find the outage probability and content delivery rate for a given cache placement. Furthermore, Tamoor-il-Hassan et al \cite{BennisISWCS15}  find the optimal station density to achieve a given hit probability, using uniform replication.


\subsection{Contributions} 

This work has the following contributions to the subject of caching at the network edge.

$\bullet$ It takes geometry explicitly into consideration for the analysis of caching policies. Specifically, it investigates a three-dimensional model (two-dimensional space and time). In this, stations have a certain spatial distribution (modelled by Point Processes) and coverage areas may overlap, allowing for multi-coverage. Furthermore, it is a dynamic model, where users with demands arrive over time at different geographic locations (Sec. \ref{Sec:3_Network}).

$\bullet$ It introduces (Sec. \ref{Sec:2_Cache}) a family of decentralised caching policies, which exploit multi-coverage, called \textit{spatial multi-LRU}. Specifically, two variations of this family are studied, namely multi-LRU-One and -All. These policies constitute an extension of the classical single-LRU, to cases where objects can be retrieved by more than one cache. The work investigates how to best choose the actions of update, insertion and eviction of content in the multiple caches and how this can be made beneficial for the performance. 

$\bullet$ The hit probability of the new policies, is analysed using the Che approximation (Sec. \ref{Sec:4_Che}). Two additional approximations made here, namely the Cache Independence Approximation (CIA) for multi-LRU-One, and the Cache Similarity Approximation (CSA) for multi-LRU-All, allow to derive simple analytical formulas for the spatial dynamic model, under IRM traffic.

$\bullet$ Verification for the Che-like approximations and further comparison of the multi-LRU policies, with other ones from the literature are provided in Sec. \ref{Sec:5_Simul} by simulations. The comparison considers policies both with distributed and with centralised implementation, that use various amount of network information. For IRM, the multi-LRU-One outperforms the -All variation. In Sec \ref{Sec:5_4_TempLoc} the policies are evaluated for traffic with temporal locality, where it is shown that multi-LRU-All can perform better than -One.


\section{Caching and its Management}
\label{Sec:2_Cache}

Caching policies can profit from the availability of system information. Such information can be related to user traffic, node positions and coverage areas, as well as the possibility for a BS to have knowledge over the cache content of its neighbours. In general, \textit{the more the available information, the higher the hit-performance}, if the management policy is adapted to it.

In general, we can group caching policies as follows.

\textbf{ (I) POQ (Policies with per-reQuest updates):} For these, updates of the cache content are done on a per-request basis and depend on whether the requested object is found or not. Information on file popularity is \textit{not available}. Neither is information over the network structure. The actions are taken \textit{locally} at each node, and are triggered by the user, in other words these policies do not require  centralised implementation. The LRU policies belong to this category.

- \textit{LRU:} it leaves in each cache the $K$ most recently demanded objects. The first position of the cache is called Most Recently Used (MRU) and the last one Least Recently Used (LRU). When a new demand arrives, there are two options. (a. \textit{Update}) The object demanded is already in the cache and the policy updates the object order by moving it to the MRU position. Or, (b. \textit{Insertion}) the object is not in the cache and it is inserted as new at the MRU position, while the object in the LRU position is \textit{evicted}. In this work we will call this policy, \textit{single-LRU}.

- \textit{q-LRU:} it is a variation of the single-LRU, with a difference in the insertion phase. When the object demanded is not in the cache, it is inserted with probability $q>0$. The eviction and order updates are the same as before.

\textbf{ (II) POP (Policies with Popularity updates):} Here, exact information over the content popularities is available. These are static policies, for which the content of caches is updated in an infrequent manner, depending on the popularity changes of the catalogue $\mathcal{F}$. The following three belong to this category.


- \textit{LFU:}  the policy statically stores in each cache the $K$ most popular contents from the set of all existing ones $\mathcal{F}$. LFU is known to provide optimal performance for a single cache under the  Independent Reference Model (IRM) . 


The next two POP policies are solutions of optimisation problems, that require a-priori knowledge of more system information additional to popularity.

- \textit{Greedy Full Information (GFI):} the policy is proposed in \cite{GolrezaeiINFOCOM12}. It assumes a-priori central knowledge of all station and user positions, their connectivity graph, and the content popularities. Using this, it greedily fills the cache memories of all stations, so that at each step of the iteration, insertion of an object at a cache is the most beneficial choice for the objective function (hit probability).

- \textit{Probabilistic Block Placement (PBP):} this policy is found in \cite{BlaGioICC15} and is similar to the GFI, with the difference that it requires less system information: the coverage number probability and the content popularities. The policy randomly assigns blocks of $K$ contents to each cache, in a way that the probability of finding a specific content somewhere in the network comes from the optimal solution of a hit maximisation problem. PBP has considerably lower computational complexity compared to GFI.


\subsection{\secit{Spatial multi-LRU}}

We propose here a novel family of distributed cache management POQ policies, that can profit from multi-coverage. We name these \textit{spatial multi-LRU} policies and are based on the single-LRU policy presented previously. The idea is that, since a user can check all the caches of covering BSs for the demanded object, and download it from any one that has it in its inventory, cache updates and object insertions can be done in a more efficient way than just applying single-LRU independently to all caches. The multi-LRU policies take into account, whether a user has found the object in \textit{any} of the covering stations, and each cache adapts its action based on this information. Most importantly, it is the user who triggers a cache's update/insertion action, and in this way she/he indirectly informs each cache about the inventory content of its neighbours. 

We propose here variations of the multi-LRU family, that differ in the number of inserted contents in the network, after a missed content demand. Differences appear also in the update phase. 

$\bullet$ \textbf{multi-LRU-One:} Action is taken only in \textit{one} cache out of $m$.  (a. Update) If the content is found in a non-empty subset of the $m$ caches, only one cache from the subset is updated. (b. Insertion) If the object is not found in \textit{any} cache, it is inserted only in one. This one can be chosen as the cache closest to the user, or a random cache, or one from some other criterion. (In this work, we will use the choice of the \textit{closest} node, to make use of the spatial independence of Poisson traffic).

$\bullet$ \textbf{multi-LRU-All:} Insertion action is taken in \textit{all} $m$ caches. (a. Update) If the content is found in a non-empty subset of the $m$ caches, all caches from this subset are updated. (b. Insertion) If the object is not found in \textit{any} cache it is inserted in all $m$.

We can also propose another variation based on q-LRU.

$\bullet$ \textbf{q-multi-LRU-All:} This variation differs from the multi-LRU-All only in the insertion phase. The object is inserted in each cache with probability $q>0$.

The motivation behind the different variations of the multi-LRU policies is the following. When a user has more than one opportunity to be served due to multi-coverage, she/he can benefit from a larger cache memory (the sum of memory sizes from covering nodes. Here we assume that the user is satisfied as long as she/he is covered, without preference over a specific station). In this setting, the optimal insertion of new content and update actions are not yet clear. If multi-LRU-One is applied, a single replica of the missed content is left down in one of the $m>1$ caches, thus favouring diversity among neighbouring caches. If multi-LRU-All is used, $m$ replicas are left down, one in each cache, thus spreading the new content over a larger geographic area (the union of $m$ covering cells), at the cost of diversity. q-multi-LRU-All is in-between the two, leaving down a smaller than $m$ number of replicas. A-priori, it is unclear which one will perform better with respect to hit probability. The performance largely depends on the type of incoming traffic. For fixed object catalogue and stationary traffic, diversity in the cache inventories can be beneficial, whereas for time-dependent traffic with varying catalogue, performance can be improved when many replicas of the same object are available, before its popularity perishes. In this work the main focus will be on spatial IRM input traffic, but a short evaluation of the policies under traffic with temporal locality will also be provided.
%
%
%
%


\section{Network Model}
\label{Sec:3_Network}

\subsection{Wireless multi-coverage}

For the analysis, the positions of transmitters coincide with the atoms from the realisation of a 2-dimensional \textit{stationary} Point Process (PP), $\Phi_b=\left\{x_i\right\}$, indexed by $i\in\mathbb{N}_+=\left\{1,2,\ldots\right\}$, with intensity $\lambda_b>0$ in $[m^{-2}]$. In this setting, the type of PP can be general, however we consider here:

- A homogeneous \textit{Poisson PP} (PPP) $\Phi_{b,P}$ with intensity measure $\mathbb{E}\left[\Phi_{b,P}(A)\right] = \lambda_b |A|$, for some area $A\subset \mathbb{R}^2$, where $|A|$ is the surface of $A$.

- A \textit{square lattice} $\Phi_{b,L} = \eta\mathbb{Z}^2+u_L$, $\mathbb{Z}=\left\{\ldots,-1,0,1,\ldots\right\}$, whose nodes constitute a square grid with edge length $\eta>0$, randomly translated by a vector $u_L$ that is uniformly distributed in $\left[0,\eta\right]^2$ (to make $\Phi_{b,L}$ stationary). Its intensity is equal to $\lambda_b=\eta^{-2}$. 

There are two different planar areas (\textit{cells}) associated with each atom (BS) $x_i$. The first one is the \textit{Voronoi cell} $\mathcal{V}(x_i)\subset\mathbb{R}^2$. Given a PP, the Voronoi tessellation divides the plane into \textit{non-overlapping} planar subsets, each one associated with a single atom. A planar point $z$ belongs to $\mathcal{V}(x_i)$, if atom $x_i$ is the closest atom of the process to $z$. In other words, $\mathcal{V}(x_i)=\left\{z\in\mathbb{R}^2: \left|z-x_i\right|\leq \left|z-x_j\right|,\ \forall x_j\in\Phi\right\}$.

The second one is the \textit{coverage cell} $\mathcal{C}_i$. Each transmitter node $x_i\in\Phi_b$ has a possibly random area $\mathcal{C}_i$ of wireless coverage associated with it. When users arrive inside the coverage cell of $x_i$ they can be served by it, by downlink transmission. In general $\mathcal{C}_i$ is different from $\mathcal{V}(x_i)$. Coverage cells can overlap, so that a user at a random location may be covered by multiple BSs, or may not be covered at all. The total coverage area from all BSs with their coverage cells is
$\Psi = \bigcup_{i\in\mathbb{N}_+}\{x_i+\mathcal{C}_i\}$ (see \cite[Ch.3]{BacBlaVol1}).

Due to stationarity of the PP $\Phi_b$, any planar location $y\in\mathbb{R}^2$ can be chosen as reference for the performance evaluation of the wireless model. This is called the \textit{typical location} $o$, and for convenience we use the Cartesian origin $(0,0)$. Because of the random realisation of the BS positions and the random choice of the reference location $o$, the number of BS cells covering $o$ is also random. 

The \textit{coverage number} $\mathcal{N}$ (as in \cite{BlaGioICC15}, \cite{KeelerBartek13}) is the number of cells that covers the typical location. It is a random variable (r.v.) that depends on the PP $\Phi_b$ and the downlink transmission scheme. It has mass function 
\begin{eqnarray}
\label{pm}
p_m := \mathbb{P}\left[\mathcal{N}=m\right], & & m=0,1,\ldots,M,
\end{eqnarray}
where $M\in\mathbb{N}_+\cup\left\{\infty\right\}$. It holds,
\begin{eqnarray}
\label{Spm}
\sum_{m=1}^{M}p_m & = & 1.
\end{eqnarray}

The choice of the coverage model determines the shape of the coverage cells and consequently the values of the coverage probabilities $p_m$. In this work the choice of $\mathcal{C}_i$ is left to be general. For the evaluation, specific models are considered. Special cases include: (1) the \textit{$\mathrm{SINR}$ Model} and (2) the \textit{$\mathrm{SNR}$ or Boolean Model}. Both models consider the coverage cell $\mathcal{C}_i$ of $x_i$, as the set of planar points for which the received signal quality from $x_i$ exceeds some threshold value $T$. The motivation is that T is a predefined signal quality, above which the user gets satisfactory Quality-of-Service. The difference between these two is that the $\mathrm{SINR}$ model refers to networks with interference (e.g. when BSs serve on the same OFDMA frequency sub-slot), whereas the $\mathrm{SNR}$ model, to networks that are noise-limited (e.g. by use of frequency reuse, neighbouring stations do not operate on the same bandwidth). For the Boolean model the $\mathcal{C}_i$ is a ball $\mathcal{B}(x_i,R_b)$ of fixed radius $R_b$ centred at $x_i$. It coincides with the $\mathrm{SNR}$ model, when no randomness of signal fading over the wireless channel is considered (or when an equivalence-type argument is used to transform the analysis of networks with random fading into equivalent ones without it, as in \cite{BartekPIMRC13}). A more detailed presentation of the different coverage models can be found in appendix \ref{app:A}.

\begin{figure*}[ht!]   
\centering  
\label{VoronoiStations}
\subfigure[Poisson Transmitters/Boolean Coverage]{
	  \centering
           \epsfig{file=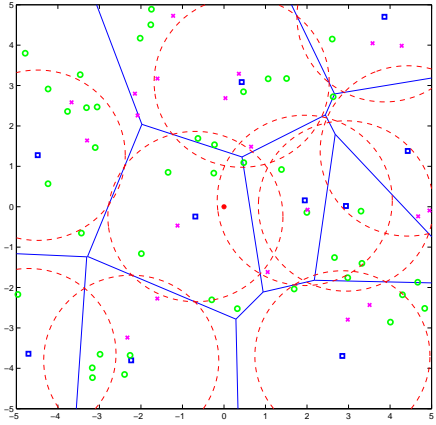, width=2.8in}
           \label{VoronoiStations:Poisson}
           }
           \subfigure[Lattice Transmitters/Boolean Coverage]{
	   \centering  
           \epsfig{file=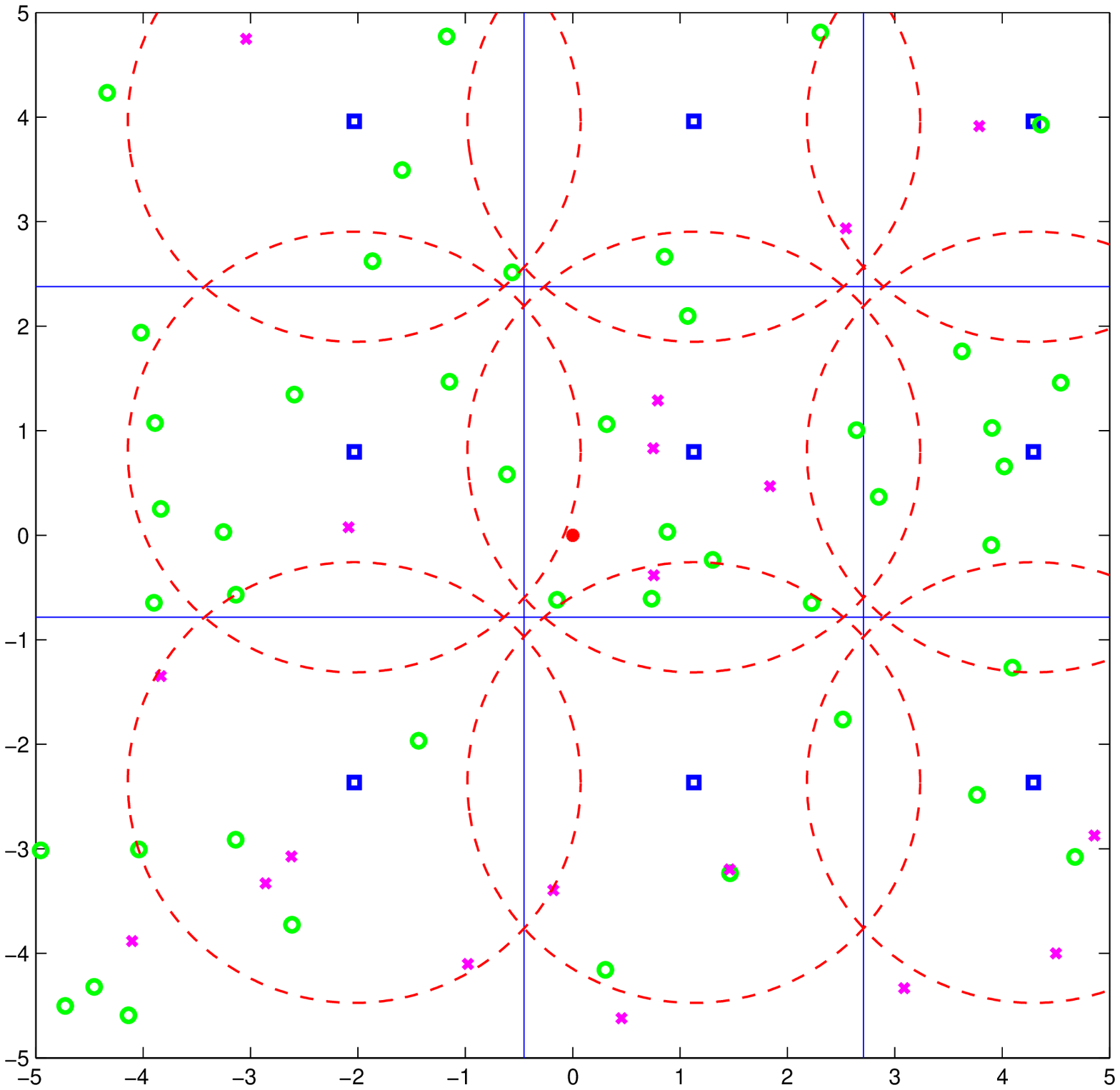, width=2.8in}
	   \label{VoronoiStations:Lattice}   
           }
\caption{A realisation of the introduced model for $t=0$ and a window of size $10\times 10$ $[m^2]$. In both subfigures, user arrivals are modelled by a PPP with $\lambda_u=0.6$ $[m^{-2} sec^{-1}]$. The users choose between two objects that have popularities $a_1=0.65$ (users with "o"), $a_2=0.35$ (users with "x"). The typical user is shown at the Cartesian origin $(0,0)$ (thicker "o"). (a) The transmitters (squares) are modelled by a PPP with $\lambda_b=0.1$ $[m^{-2}]$. (b) The transmitters (squares) are modelled by a Square Lattice PP with $\eta = \lambda_b^{-1/2}=1/\sqrt{0.1}$ $[m]$. In both figures, we assume the Boolean model for coverage, with $R_b=2\eta/3$ $[m]$.}
\end{figure*}

\subsection{Spatial IRM Traffic model} 

Each user served from the network is assumed to arrive independently at some planar location, stay there during service and then leave. We model the users by a \textit{homogeneous space-time} PPP in $\mathbb{R}^2\times \mathbb{R}$, $\Phi_u=\left\{\left(\psi_i,t_i\right)\right\}$, where $\psi_i$ takes values on the Euclidean plane, and the time $t_i$ of arrival occurs at some point on the infinite time axis. The PPP intensity is $\lambda_u>0$ in $[m^{-2}sec^{-1}]$. Service time is considered fixed and equal to unity but 
it will not play any role in the analysis. 
Given a planar area $A$, the arrival rate of users in this area is equal to $\lambda_u|A|$ in $[sec^{-1}]$. The time between two consecutive arrivals in $A$ is exponentially distributed with mean $(\lambda_u|A|)^{-1}$ $[sec]$ and all users within the area take their positions independently and uniformly. 

Each user arrives with a request for a specific data object. In this work, we follow the so-called Independent Reference Model (IRM)  \cite{FaPrSIAM78}, according to which (i) the catalogue of available objects, denoted by $\mathcal{F}$, has finite size $F$. (ii) The probability $a_j$ that a user requests object $c_j\in\mathcal{F}$ (i.e. the object \textit{popularity}) is constant (does not vary over time), known, and independent of all past requests. In this way, the sequence of generated requests in space and time is i.i.d.. We additionally consider that all objects have the same size, normalised to 1. Cases of unequal size will not be treated in this work, but we can always assume that each file can be divided into chunks of equal size, so the same analysis can still be applied. Objects in $\mathcal{F}$ are ordered by popularity: $c_1$ is the most popular, $c_2$ the second most popular and so on. The popularity of $c_j$ is $a_j>0$, and to be consistent with the ordering, we also have $a_1\geq a_2\geq\ldots \geq a_F$. For every popularity distribution it obviously holds,
\begin{eqnarray}
\label{sumAJ}
\sum_{j=1}^{F}a_j  & = & 1.
\end{eqnarray}

Without loss of generality, we will consider (especially in the simulations) that the distribution has a Zipf probability mass function, although the analysis holds for general $\left\{a_j\right\}$. This is motivated by traffic measurements showing that data-object popularity in the WWW follows a power law \cite{HuberWWW99}, \cite{Newman05}. In such case, the probability that a user asks for $c_j$ is equal to 
$a_j  =  D^{-1}j^{-\gamma}$, $j=1,\ldots,F$.
Here, $\gamma$ is the Zipf exponent, often chosen as $\gamma<1$, so that $a_1/a_2=2^{\gamma}<2$. The normalisation factor is equal to $D:=\sum_{j=1}^{F} j^{-\gamma}$. 

In order to incorporate the object request process in the analysis, we relate to each atom $\left(\psi_i,t_i\right)$ of the user process $\Phi_u$ a \textit{mark} $v_i$. Each mark $v_i$ is an independent realisation of the random variable $V$ (and independent of location and time) taking as values the indices of the objects $c_j\in\mathcal{F}$, and has distribution $\left\{a_j\right\}$. In this way, we define the \textit{iid marked PPP} $\hat{\Phi}_u=\left(\psi_i,t_i,v_i\right)$ on $\mathbb{R}^2\times\mathbb{R}\times\mathcal{F}$. A consequence of the independent marking, is that the users that request object $c_j\in\mathcal{F}$ form a homogeneous space-time PPP with intensity $a_j\lambda_u$ $[m^{-2}sec^{-1}]$, which results from an independent thinning of $\Phi_u$.

The way we have modelled user traffic (using IRM) ignores temporal and/or spatial correlations in the sequence of user requests, since it assumes independence in all dimensions. In reality however, when an object is requested by a user, it is more likely to be requested again at some near future in a neighbouring location. This is called time-locality \cite{TraversoTranMult15}, \cite{OlmosTEMPO14} and space-locality \cite{BroSceWa12}. The presented PP model has the flexibility to be adapted to such traffic behaviour. We will not give much details about this type of traffic in this paper (the reader is referred to the related references). Some first simulations for the policies under study are however provided in Sec. \ref{Sec:5_4_TempLoc}. Further research on this is the subject of our ongoing work.

We consider the case where a cache memory of size $K\geq 1$ is installed and available on each transmitter node $x_i$ of $\Phi_b$. The memory inventory of node $x_i$ at time $t$ is denoted by $\Xi_i(t)$ and is a (possibly varying over time) subset of $\mathcal{F}$, with number of elements not greater than $K\geq 1$.


\begin{table}[t!]
\caption{Symbols}
\centering
\begin{tabular}{|c | l |}
\hline
$\Phi_b$ 							& Point Process of transmission nodes $\left\{x_i\right\}$\\
$\Phi_{b,P}$, $\Phi_{b,L}$ 			& Poisson and Lattice position of $\left\{x_i\right\}$\\
$\Phi_u$ 							& Point Process (Poisson) of users $\left\{(\psi_i,t_i)\right\}$\\
$\lambda_b$ 						& intensity of transmission nodes [$m^{-2}$]\\
$\lambda_u$ 						& intensity of users [$m^{-2}sec^{-1}$]\\
$A$ 								& planar area\\
$\mathcal{V}(x_i)$					& Voronoi cell of node $x_i$\\
$\mathcal{C}_i$					& coverage cell of node $x_i$\\
$R_b$ 							& radius of coverage\\
$p_m$							& probability of coverage by m nodes\\
$\mathcal{F}$						& object catalogue of size $F$\\
$a_j$							& popularity of object $c_j\in\mathcal{F}$\\
$\hat{\Phi}_u$ 						& PPP of users marked by object $\left\{(\psi_i,t_i,c_i)\right\}$\\
$o$, $u_o$ 						& Typical location and typical user\\
$K$ 								& size of cache memory\\
$\Xi_i(t)$							& inventory of cache on BS $x_i$ at time $t$\\
\hline
\end{tabular}
\label{TofS}
\end{table}


The network performance is evaluated at the \textit{typical user} $u_o$, who - due to stationarity of the PPP - will be representative of any user of the process. We suppose that this user appears at the Cartesian origin $(0,0)$, at time $t_o=0$. 
In this way, the \textit{typical user} coincides with the \textit{typical location}  $o$ of the process $\Phi_b$ at time $t=0$. 
The model described so far is illustrated in Fig.\ref{VoronoiStations:Poisson} for the case of Poisson placement of transmitters $\Phi_{b,P}$ with Poisson arrivals $\Phi_u$, and in Fig.\ref{VoronoiStations:Lattice} for the case of a square lattice $\Phi_{b,L}$ with Poisson arrivals $\Phi_u$. In both cases the Boolean coverage model is considered and the typical user is shown at $(0,0)$ for $t=0$. In this realisation, the typical user is covered by two cells in the PPP case and by a single one in the Lattice case. We also provide the reader with a list of symbols in Table \ref{TofS}.

As mentioned already, the performance measure of the caching policies is the hit probability. We can already provide an upper bound for any POP policy (and consequently any POQ, since these use less information). The bound requires knowledge over the content popularity and coverage number, like the PBP. (Due to stationarity of $\Phi_b$, it is also a bound for GFI). Specifically, the hit probability of a user covered by $m$ cells is maximised if these $m$ inventories have distinct entries, so that the user has the maximum choice. The $mK$ objects installed should be the most popular ones from the set $\mathcal{F}$. So the upper bound is equal to,

\begin{eqnarray}
\label{UBhit}
P_{hit} \leq \sum_{m=1}^M p_m \sum_{j=1}^F a_j \mathbbm{1}_{\left\{1\leq j\leq mK\right\}} =  \sum_{m=1}^M p_m \sum_{j=1}^{mK} a_j. 
\end{eqnarray}

\section{Che-Like Approximations}
\label{Sec:4_Che}

\subsection{Single cache}
The mathematical analysis of LRU policies is complicated, due to the different inter-arrival times for different content and the update(eviction)/insertion policy. However, Che et al provided in 2002 \cite{CheApprox02} an analysis and a simple approximation for the single-LRU cache, which results in an analytical formula for the hit probability $P_{hit}$ with excellent fit to simulations. In the following, we explain in short the idea and, after, apply it to the multi-LRU policies.

The approximation is based on the so-called \textit{characteristic time} $T_C$. Given a cache of size $K$ under single-LRU replacement, if at time $t=0$ an arrival of object $c_j$ occurs, then this will be positioned at the MRU place, either due to a. Update, or due to b. Insertion. This object is removed from the cache if at least $K$ different objects arrive, before a new demand for object $c_j$ at time $\tau_j>0$. The reason is that, each arrival of a new object, moves $c_j$ one position away from the MRU and closer to the LRU. 
The Che approach approximates the eviction time of an object by a deterministic quantity, equal for all objects to the characteristic time $T_C$. This is found by solving
\begin{eqnarray}
\label{Cheapprox}
\sum_{i=1}^F \mathbb{P}\left(\tau_i<T_C\right) = K & & (Che\ approximation),
\end{eqnarray}
using a fixed point procedure, where $\tau_i$ is the first arrival time of object $c_i$, $i\neq j$, after $t=0$. The summation in (\ref{Cheapprox}) is taken over the entire $\mathcal{F}$, which is also part of the approximation. It works well for a large number $F$ of objects, each one of which having a small portion of the popularity. For IRM traffic, the inter-arrival times are exponentially distributed, hence for an area $A$ covered by a single cache, $\mathbb{P}\left(\tau_i<T_C\right) = 1-e^{-\lambda_u |A|a_i T_C}$. The time-average probability that an object $c_j$ is in the cache is
\begin{eqnarray}
\label{Pin}
\mathbb{P}\left(c_j\in\Xi\right) = \mathbb{P}\left(\tau_j<T_C\right) \stackrel{IRM}{=} 1-e^{-\lambda_u |A|a_j T_C} = P_{hit}(j).
\end{eqnarray}
The fact that, for IRM traffic $\mathbb{P}\left(c_j\in\Xi\right) = P_{hit}(j)$, is due to the PASTA property of Poisson arrivals. Finally, the approximation for the total hit probability is, 
\begin{eqnarray}
\label{PhitChe}
P_{hit} & = & \sum_{j=1}^F a_j P_{hit}(j).
\end{eqnarray}

%
\subsection{General Approximation for multi-LRU}
\label{GenChe}


We will use the approach of Che for the single-LRU, to derive here similar approximations of the multi-LRU cache management policies, for the network model described in the previous section. (To provide more intuition on this general approach, a similar analysis for a network with only two caches is given in appendix \ref{app:B}).

Consider an arrival of user $u_o$ at the Cartesian origin $\psi_o=(0,0)$, at time $t_o=0$, who requests for object $c_j$. This is the \textit{typical user}, who is covered by a number $m_o\geq 0$ of BSs, a realisation of the r.v. $\mathcal{N}_o$ with mass function $\left\{p_m\right\}$. A common characteristic time $T_C$ is assumed for all caches of the network, due to stationarity of all processes. We focus on the cache of a specific $x_i$ among the $m_o$ covering BSs, for which definitely $o\in\mathcal{C}_i$. The probability that user $u_o$ finds the requested content in the cache of $x_i$, is calculated using the following arguments:  The previous user requesting for the same object $c_j$ arrived in an area $\mathcal{S}_{-1}$ (that varies depending on the type of multi-LRU policy) and is covered by $x_i$ definitely (otherwise the user will not influence $\Xi_i$) and possibly some other stations, the total number of which is $\tilde{m}$ (the realisation of another r.v. $\mathcal{N}_{-1}$). Since we know that $u_{-1}$ is at least covered by one station (the $x_i$), the distribution of $\mathcal{N}_{-1}$ has mass function 
\begin{eqnarray}
\label{pmac}
\tilde{p}_{\tilde{m}} = \frac{p_{\tilde{m}}}{1-p_0}, & & \tilde{m}=1,\ldots,M.
\end{eqnarray}

Suppose this user arrived at $t_{-1}^-\in\left|t_o-T_C,t_o\right|$, i.e. within the characteristic time ($t^-$ is the time right before $t$). Then 
the object is found in $\Xi_i(t_o^-)$ at $t_o^-$, if (i) either the object was in $\Xi_i(t_{-1}^-)$ and an update was triggered by $u_{-1}$, or (ii)  the object was not cached in any of the $\tilde{m}$ stations and an insertion in the inventory $\Xi_i$ was triggered. If $m_o>0$ (otherwise, the user is not covered), we write for $ i\in\left\{1,\ldots,m_o\right\}$
\begin{eqnarray}
\label{Phitgen1}
P_{hit,i}(u_o) & = & \mathbb{P}\left(u_{-1}\in(\mathcal{S}_{-1},|t_o-t_{-1}|<T_C,j)\right)\cdot\nonumber\\
& & \cdot \left[\mathbb{P}(c_j\in\Xi_i(t_{-1}^-))+\mathbb{P}(\bigcap_{\ell=1}^{\tilde{m}}\left\{c_j\notin\Xi_\ell(t_{-1}^-)\right\})\right].\nonumber
\end{eqnarray}
For IRM traffic with PASTA, $\mathbb{P}(c_j\in\Xi_i(t_{-1}^-)) = P_{hit,i}(u_{-1})$, and is also independent of the time $t$ and user position $\psi$, hence we can simply write $P_{hit,i}(j)$. Substitution in the above equation gives,
\begin{eqnarray}
\label{Phitgen2}
P_{hit,i}(j) & = & \mathbb{P}\left(u_{-1}\in(\mathcal{S}_{-1},|t_o-t_{-1}|<T_C,j)\right)\cdot\nonumber\\
& & \cdot \left[P_{hit,i}(j) +\sum_{\tilde{m}=1}^{M} \frac{p_{\tilde{m}}}{1-p_o}\mathbb{P}(\bigcap_{\ell=1}^{\tilde{m}}\left\{c_j\notin\Xi_\ell\right\})\right].\nonumber\\
\end{eqnarray}
Solving the above over $P_{hit,i}(j)$ provides an expression for the hit probability of object $c_j$ at the cache of node $x_i$. To find the characteristic time $T_C$ we solve the equation, (in the IRM case)
\begin{eqnarray}
\label{TCgen}
\sum_{j=1}^F P_{hit,i}(j) = K, & & i\in\left\{1,\ldots,m_o\right\}.
\end{eqnarray}
Finally, the total hit probability is equal to,
\begin{eqnarray}
\label{PhitTOTgen}
P_{hit} 
& = & \sum_{j=1}^Fa_j\sum_{m_o=0}^{M}p_{m_o}\left(1-\mathbb{P}(\bigcap_{\ell=1}^{m_o}\left\{c_j\notin\Xi_\ell\right\})\right).
\end{eqnarray}
We note that $\mathbb{P}(\bigcap_{\ell=1}^{0}\left\{c_j\notin\Xi_\ell\right\}) = 1$, for $m_o=0$, in which case, the user surely misses the content.

The main difficulty when dealing with the general case, is that the hit probability of one cache depends on the hit probability of its neighbours and the neighbours of its neighbours. This is because the coverage area of each node has many sub-areas of multi-coverage by different BS subsets, which makes analysis neither easy, nor exact. 

\subsubsection{multi-LRU-One (Che with CIA)} 
Only the users falling in the Voronoi cell of a node can trigger an action of a. Update or b. Insertion at the cache of that node as long as they are covered. Then $\mathcal{S}_o=\mathcal{S}_{-1}=\mathcal{V}(x_i)$ in (\ref{Phitgen1}). The coverage cell can be smaller than the Voronoi cell, in which case, only the users falling in the intersection of the two, trigger cache actions. To avoid dealing with these special cases, we consider coverage cells which fully cover the related Voronoi cells, that is $|\mathcal{C}_i|>|\mathcal{V}_i|$. $\forall i$.


There are the unknown probabilities $\mathbb{P}(\bigcap_{\ell=1}^{\tilde{m}}\left\{c_j\notin\Xi_\ell\right\})$ and  $\mathbb{P}(\bigcap_{\ell=1}^{m_o}\left\{c_j\notin\Xi_\ell\right\})$ that need to be calculated. Instead of directly trying to find a solution, we use a \textit{Cache Independence Approximation} (CIA). Based on this, each cache performs single-LRU for the users that arrive within its Voronoi cell. The idea is that, since only the users in the Voronoi cell change the inventory of the related cache, the influence of the neighbouring stations' traffic on the inventory of $x_i$ should be small. Then in (\ref{Phitgen1}) we forget the rest $\tilde{m}-1$ nodes and we replace 
\begin{eqnarray}
\label{CIA1}
\mathbb{P}(\bigcap_{\ell=1}^{\tilde{m}}\left\{c_j\notin\Xi_\ell\right\}) \approx \mathbb{P}(c_j\notin\Xi_i), & (CIA_1).
\end{eqnarray}
Furthermore, the independence due to the CIA, has the result that, when the user is covered by $m_o$ stations, her/his hit probability is simply the product of hit probabilities of all these stations. The fact that the Voronoi cells of different stations do not overlap is further in favour of the approximation. Then, in (\ref{PhitTOTgen})
\begin{eqnarray}
\label{CIA2}
\mathbb{P}(\bigcap_{\ell=1}^{m_o}\left\{c_j\notin\Xi_\ell\right\}) \approx (\mathbb{P}(c_j\notin\Xi_i))^{m_o}, & (CIA_2).
\end{eqnarray}
From the above, the hit probability of each object in $\Xi_i$ is,
\begin{eqnarray}
\label{Phitgen2one}
P_{hit,i}(j) & = & \mathbb{P}\left(u_{-1}\in(\mathcal{S}_{-1}\in\mathcal{V}(x_i),|t_o-t_{-1}|<T_C,j)\right)\cdot\nonumber\\
& &  \cdot\left[P_{hit,i}(j) +\mathbb{P}(c_j\notin\Xi_i)\right]\nonumber\\
& \stackrel{IRM}{=}& 1- e^{-a_j\lambda_u|\mathcal{V}|T_C}, \ i\in\left\{1,\ldots,m_o\right\}.
\end{eqnarray}
We used the fact that for IRM $P_{hit,i}(j) = 1- \mathbb{P}(c_j\notin\Xi_i)$.
The characteristic time is found by solving the equation
\begin{eqnarray}
\label{TConegen}
\sum_{j=1}^F (1- e^{-a_j\lambda_u|\mathcal{V}|T_C}) =K.
\end{eqnarray}
The total hit probability, based on CIA, is, 
\begin{eqnarray}
\label{PhitTOTgenone}
P_{hit} & = & \sum_{j=1}^Fa_j\sum_{m_o=0}^{M}p_{m_o}\left(1-\mathbb{P}(c_j\notin\Xi_i)^{m_o}\right)\nonumber\\
& \stackrel{(\ref{Phitgen2one})}{=} & \sum_{j=1}^Fa_j\sum_{m_o=0}^{M}p_{m_o}\left(1-e^{-a_j\lambda_um_o|\mathcal{V}|T_C}\right).
\end{eqnarray}
Special case: For the PPP model of node positions, it is known \cite{BacBlaVol1} that the average size of a Voronoi cell is equal to $|\mathcal{V}|=\lambda_b^{-1}$. In the Boolean coverage model, $|\mathcal{C}|=\pi R_b^2$. 


\subsubsection{multi-LRU-All (Che with CSA)} 
In this case, users falling on any point inside the coverage cell of $x_i$ can trigger an action of update and insertion at its cache inventory $\Xi_i$. This means that $\mathcal{S}_{o}=\mathcal{S}_{-1}=\mathcal{C}_i$, for the hit probability expression in (\ref{Phitgen1}). 

Again, the unknown probabilities $\mathbb{P}(\bigcap_{\ell=1}^{\tilde{m}}\left\{c_j\notin\Xi_\ell\right\})$ and $\mathbb{P}(\bigcap_{\ell=1}^{m_o}\left\{c_j\notin\Xi_\ell\right\})$ need to be calculated. 
In this case, we use a different approximation, the \textit{Cache Similarity Approximation} (CSA), which states that inventories of neighbouring caches have the same content. This is motivated by the fact that new content is simultaneously installed in all caches of nodes covering a user, when the user triggers insertion. The approximation is better, the larger the cache size $K$, because for large memories it takes more time for an object to be evicted after its insertion and similar content stays in all inventories. Then in (\ref{Phitgen1}),
\begin{eqnarray}
\label{CSA1}
\mathbb{P}(\bigcap_{\ell=1}^{\tilde{m}}\left\{c_j\notin\Xi_\ell \right\}) \approx  \mathbb{P}(\left\{c_j\notin\Xi_i\right\}), & (CSA_1).
\end{eqnarray}
Interestingly, $CSA_1$ and $CIA_1$ give the same expression. However, in multi-LRU-All, we do not assume independence, but rather similarity. Then, since neighbouring caches have the same content, the total miss probability when a set of $m_o$ stations cover user $u_o$ is equal to the probability that no user with the same demand arrives within the total area of coverage during the characteristic time $T_C$ (otherwise the content is definitely in all caches, either because of a. Update or b. Insertion. Then, for IRM traffic,
\begin{eqnarray}
\label{Pmissmoall}
\mathbb{P}(\bigcap_{\ell=1}^{m_o}\left\{c_j\notin\Xi_\ell\right\}) \approx e^{-a_j\lambda_u|\mathcal{A}_{m_o}|T_C}, & (CSA_2).
\end{eqnarray}
In the above, the total area of coverage from the $m_o$ stations is denoted by $\mathcal{A}_{m_o}$ and its surface is equal to,
\begin{eqnarray}
\label{SurfaceM}
\left|\mathcal{A}_{m_o}\right| & = & \left|\bigcup_{\ell=1}^{m_o}\mathcal{C}_i\right|, \ \  m_o=0,\ldots,M.
\end{eqnarray}
It holds $|\mathcal{A}_0|=0$, for $m_o=0$. For the Boolean model $|\mathcal{A}_{1}| = \left|\mathcal{C}_1\right| = \pi R_b^2$, while the surface of $\mathcal{A}_{m_o}$ is a superposition of $m_o$ overlapping discs with equal radius $R_b$.

The hit probability of each object in $\Xi_i$ is found by using CSA in (\ref{Phitgen1}), and we get
\begin{eqnarray}
\label{Phitgen2all}
P_{hit,i}(j) & = & \mathbb{P}\left(u_{-1}\in(\mathcal{S}_{-1}\in\mathcal{C}_i,|t_o-t_{-1}|<T_C,j)\right)\cdot\nonumber\\
& &  \cdot \left[P_{hit,i}(j) +\mathbb{P}(\left\{c_j\notin\Xi_i\right\})\right]\nonumber\\
& \stackrel{IRM}{=}& 1- e^{-a_j\lambda_u|\mathcal{C}|T_C}.
\end{eqnarray}
We used the fact that for IRM $P_{hit,i}(j) = 1- \mathbb{P}(c_j\notin\Xi_i)$. For the characteristic time, we solve the equation
\begin{eqnarray}
\label{TCallgen}
\sum_{j=1}^F (1- e^{-a_j\lambda_u|\mathcal{C}|T_C}) & = & K.
\end{eqnarray}
The total hit probability, based on CSA, is
\begin{eqnarray}
\label{PhitTOTgenall}
P_{hit} & \stackrel{(\ref{Pmissmoall})}{=} & \sum_{j=1}^Fa_j\sum_{m_o=0}^{M}p_{m_o}\left(1-e^{-a_j\lambda_u|\mathcal{A}_{m_o}|T_C} \right).
\end{eqnarray}
The only difficulty in calculating the approximate hit probability for multi-LRU-All with the above formulas, is to obtain exact values for the total surface $|\mathcal{A}_{m_o}|$.

(We refer, again, the reader to appendix \ref{app:B} for the two-cache network example.)
%
%
 \section{Simulation and Comparison}
 \label{Sec:5_Simul}

We have performed extended simulations in order to verify the Che-like approximations and also to evaluate and compare the proposed multi-LRU policies with other ones from the literature. The comparison is based on the hit-probability performance measure and assumes IRM traffic, except in Sec. \ref{Sec:5_4_TempLoc} where traffic with temporal locality is studied. 

\subsection{Simulation setup}

BSs are placed within a rectangular window of size $A \times B = 12\times 12$ $[km^2]$. After choosing the BS intensity $\lambda_b=0.5$ $[km^{-2}]$, their positions are chosen based on the type of network we want to analyse (PPP or Lattice). For PPP, a Poisson number of stations is simulated in each realisation and their positions are set uniformly inside the window. In the case of a Lattice network, the stations are put on a square grid with distance $\eta=1/\sqrt{\lambda_b} = 1.4142$ $[km]$ from each other. In both types of networks, the average Voronoi size $|\mathcal{V}|=\lambda_b^{-1}$ (see \cite{BacBlaVol1}).

We evaluate a Boolean coverage model so that every station covers a disc of radius $R_b\in\left[0.5, 3\right]$ $[km]$ with surface $|\mathcal{C}|=\pi R_b^2$. The larger the radius the more the multi-coverage effects. The magnitude of coverage overlap can be described by the expected number of BSs covering a planar point, $\overline{N_{BS}}=\mathbb{E}[{\text{Number of covering stations}}] = \sum\limits_{m=1}^{M}m p_m$, where the $p_m$ are the coverage number probabilities for $m$ stations, whose values depend on the node placement and coverage model (PPP or Lattice). The maximum number of covering stations is chosen $M=50$. For the Boolean PPP case, the probabilities $\left\{p_m\right\}$ correspond to a Poisson r.v. with parameter $\nu:=\lambda_b\pi R_b^2$ (see (\ref{pBool})). For the Boolean Lattice case, these are found by Monte Carlo simulations. Given the intensity, $\lambda_b = 0.5$, there is a mapping from the Boolean radius $R_b$ to the number $\overline{N_{BS}}$, some values of which are given in Table \ref{tab:Radius_Nexp}.
\begin{table}[h]

\caption{$R_b$  to $\overline{N_{BS}}$ mapping for Boolean PPP and Lattice ($\lambda_b=0.5$ $km^{-2}$).}
\centering
\begin{tabular}{|c|c|c|}

\hline
Radius ($R_b$) $[km]$	& PPP ($\overline{N_{BS}}$) & 	Lattice ($\overline{N_{BS}}$) \\
\hline
\hline
0.8 	& 1 & 1.06\\
\hline
1.13 & 2 & 2.12\\
\hline
1.38 & 3 & 3.22\\
\hline
1.60 & 4 & 4.21\\
\hline
1.78 & 5 & 5.32\\
\hline
1.95 & 6 & 6.42\\
\hline
2.11 & 7 & 7.43\\
\hline
2.26 & 8 & 8.44\\
\hline
\end{tabular}
\label{tab:Radius_Nexp}
\end{table}

Following the spatial IRM traffic model for the request arrivals, we consider a homogeneous space-time PPP with intensity $\lambda_u=0.023$ $[m^{-2}sec^{-1}]$, which is approximately equal to 80 $[m^{-2}/hour^{-1}]$ requests - a reasonable value for a busy corner in a city. Based on the model, the expected number of requests within the entire window $A\times B$ in a time interval of $T_s=1$ $[month]$ (30 days) is equal to $\lambda_{TS} = \lambda_uABT_s = 0.023\cdot 12^2 \cdot 30\cdot 24\cdot 3600$ $=$ $8.622\cdot 10^6$. For each realisation of a BS deployment we produce a number of total requests from a Poisson distribution with parameter $\lambda_{TS}$. These requests are uniformly positioned within the interval $[0,T_{s}]$. Each request is given two marks. The first one is its location on the window. The location marks are i.i.d. vectors having entries the $(x,y)$ coordinates of the request, where the latter are chosen uniformly within the interval $[0,A]$ and $[0,B]$, respectively.

The second mark of each request is the content demand taken from a catalogue of size $F=10,000$ objects. The popularities of these objects follow a Zipf distribution with parameter $\gamma=0.78$  (unless otherwise stated). A cache memory of capacity $K$ objects is considered available on each BS. The size $K$ is defined as a proportion of the catalogue size, i.e. $K=\alpha F$. In the evaluation/simulations $\alpha$ can take values $0.01$, $0.02$ or $0.05$. This means that $1\%$, $2\%$ or $5\%$ of the catalogue size can be cached in the memory of each BS. The $\alpha$ parameter is called the \textit{Memory-to-Catalogue-size Ratio}.


When a user is covered by a station with the requested content in memory, the demand is considered a hit. At the end of the simulation of a large number of realisations for the BS and request point processes (this number chosen over 10,000) the total hit probability is approximated by the frequency of hits (number of hits over number of requests).

Obviously, the window size plays an important role, due to edge effects. For requests near the edges, there could be BSs outside the simulation window that would provide coverage, but due to the finiteness, their influence is omitted. To diminish this effect, which results in reduced hit probability, stations are added in an outer rectangular window $(A+R_b) \times (B+R_b)$, while users are considered to arrive only within the original main window.

\subsection{Verification of the approximations}
To verify the validity of the proposed approximations, we compare the results of the general model in the previous section with the hit probability from simulations, for the Boolean PPP case. 
For the memory size $K$, we consider two cases, (a) $\alpha=0.05\Rightarrow K=0.05 F$, hence $K = 500$ objects, and (b) $\alpha=0.2\Rightarrow K=0.2 F$, hence $K=2000$ objects. 

\begin{figure*}[t!]   
\centering  
\label{multiLRUoneallVerif}
\subfigure[multi-LRU-One]{
	  \centering
           	\epsfig{file=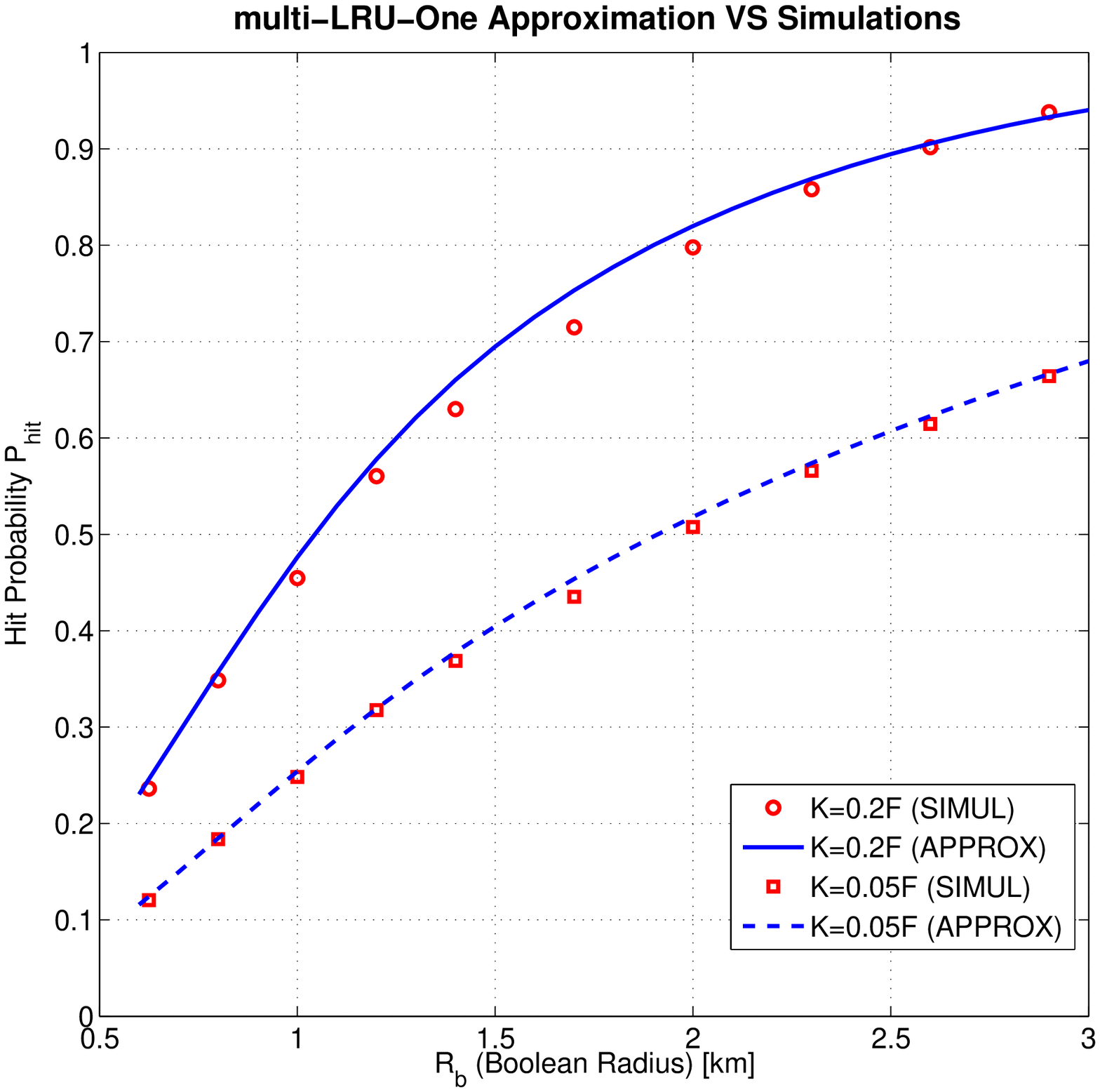, width=3in}
           \label{Verif:ONE}
           }
           \subfigure[multi-LRU-All]{
	   \centering  
           	 \epsfig{file=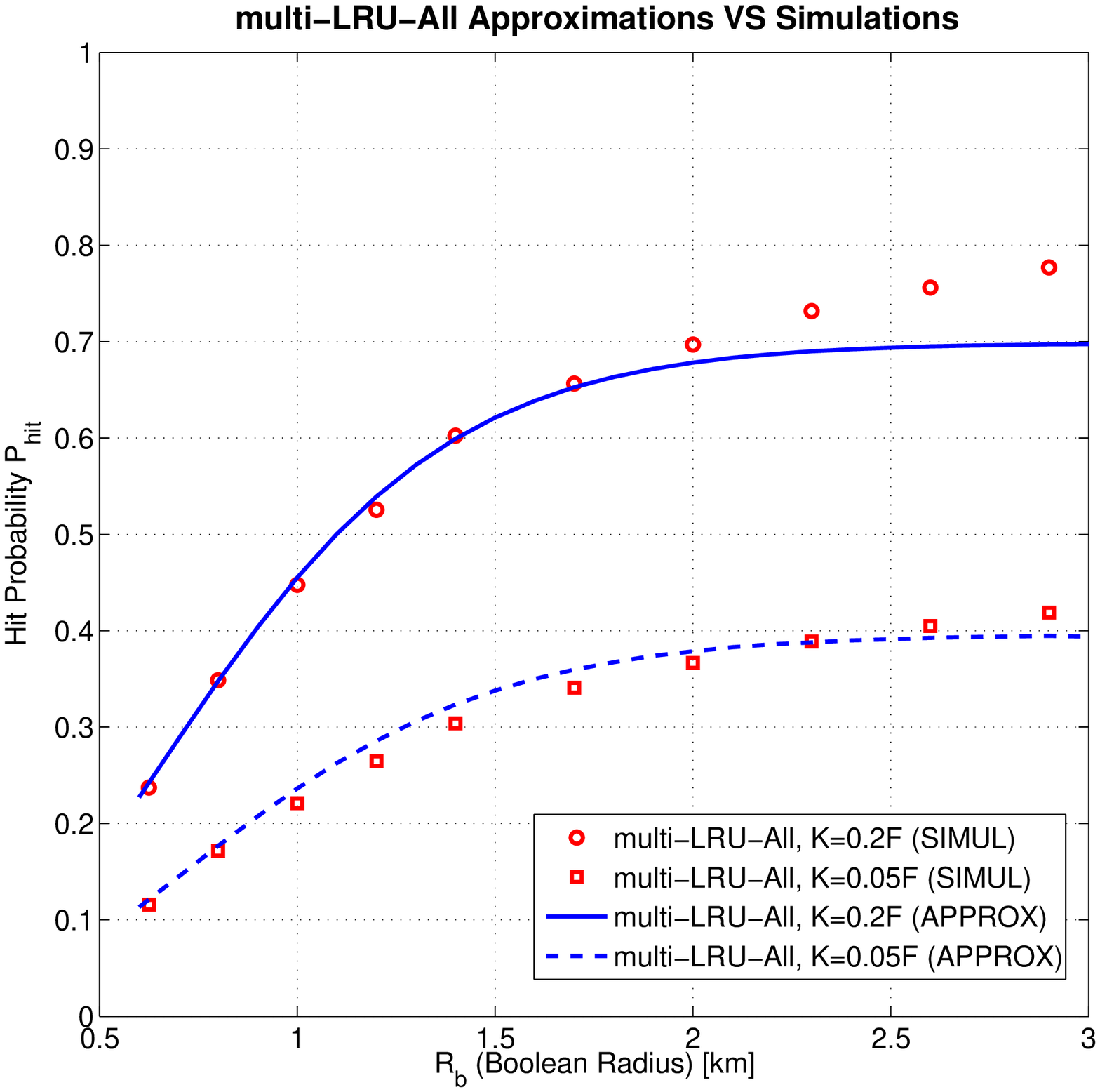, width=3in}
	   \label{Verif:ALL}   
           }
         \caption{Verification of the Che-Like approximations for the two multi-LRU policies.}
\end{figure*}

$\bullet$ \textbf{multi-LRU-One}: The total hit probability is evaluated numerically using (\ref{PhitTOTgenone}). The characteristic time per cache is found by solving (\ref{TConegen}) by a fixed point method, where the individual hit probability of each object is given in (\ref{Phitgen2one}). To guarantee that $|\mathcal{C}|>|\mathcal{V}|$, we need that $\pi R_b^2>\lambda_b^{-1}$ $\Rightarrow$ $R_b>(\pi\lambda_b)^{-{1/2}}=0.4$. Since $R_b>0.6$ in the evaluation, the condition should be satisfied. The comparison between approximate hit probability and simulations are shown in Fig. \ref{Verif:ONE}. The curves exhibit a very good match. The evaluation shows that the independence approximation (CIA) works very well in this general model with PPs. 

$\bullet$ \textbf{multi-LRU-All}: The total hit probability is evaluated numerically using (\ref{PhitTOTgenall}). The characteristic time per cache is found by solving (\ref{TCallgen}) using a fixed point method, where the individual hit probability of each object is given in (\ref{Phitgen2all}). 

We provide a method to estimate the surfaces $|\mathcal{A}_{m_o}|$, $m_o=1,\ldots,M$ for the Boolean/PPP case: A user $u_o$ has a distance $R_{d,i}$ from each one of the $m_o$ nodes $x_i$ that cover her/him. These distances are realisations of a random variable, whose expected value can be found equal to $\mathbb{E}[R_d]= 2R_b/3$, i.e. the user lies in expectation at $2R_b/3$ away from the center of each covering disc. Then we have:

1) The coverage cell size (for the Boolean model) is the disc surface, equal to $|\mathcal{A}_1|=|\mathcal{C}|=\pi R_b^2$.

2) When $M\rightarrow\infty$, a disc having center the user $u_o$ and radius $R_M$ $=$ $R_b+\mathbb{E}[R_d]$ $=$ $5R_b/3$ is (due to randomness of node positions) fully covered. So $|\mathcal{A}_M|= |\mathcal{C}|(5/3)^2$.

3) For intermediate cases $1<m_o<M$, the surface should be somewhere between the two extremes, and obviously the surface $|\mathcal{A}_{m_o}|$ should be monotone increasing with $m_o$. We also expect that  for low $m_o$, the total area $|\mathcal{A}_M|$ will be filling fast, whereas for larger ones, the change in surface should be small. For this we can use a function with exponential decrease for large $m_o$, such as
\begin{eqnarray}
\label{Amo}
|\mathcal{A}_{m_o}| = |\mathcal{A}_M|(1-e^{-m_o\rho}), & \rho = -\ln(1-\frac{\mathcal{A}_1}{\mathcal{A}_M}).
\end{eqnarray}
The comparison between approximate hit probability and simulations are shown in Fig. \ref{Verif:ALL}. The approximation and simulation curves seem to closely follow one another. For large values of the radius, the approximation curves seem to diverge from the simulations. This should be less a failure of the CSA approximation (which has been shown to be accurate for the two-cache network in appendix \ref{app:B}), but more possibly a failure of the above method to approximate the surfaces $|\mathcal{A}_{m_o}|$. More accurate values of $|\mathcal{A}_{m_o}|$ should exhibit a better fit.

%
\subsection{Comparison of policies}

\subsubsection{Hit Probability versus Coverage Number} 

\begin{figure}[h!]   
	\centering  
    \epsfig{file=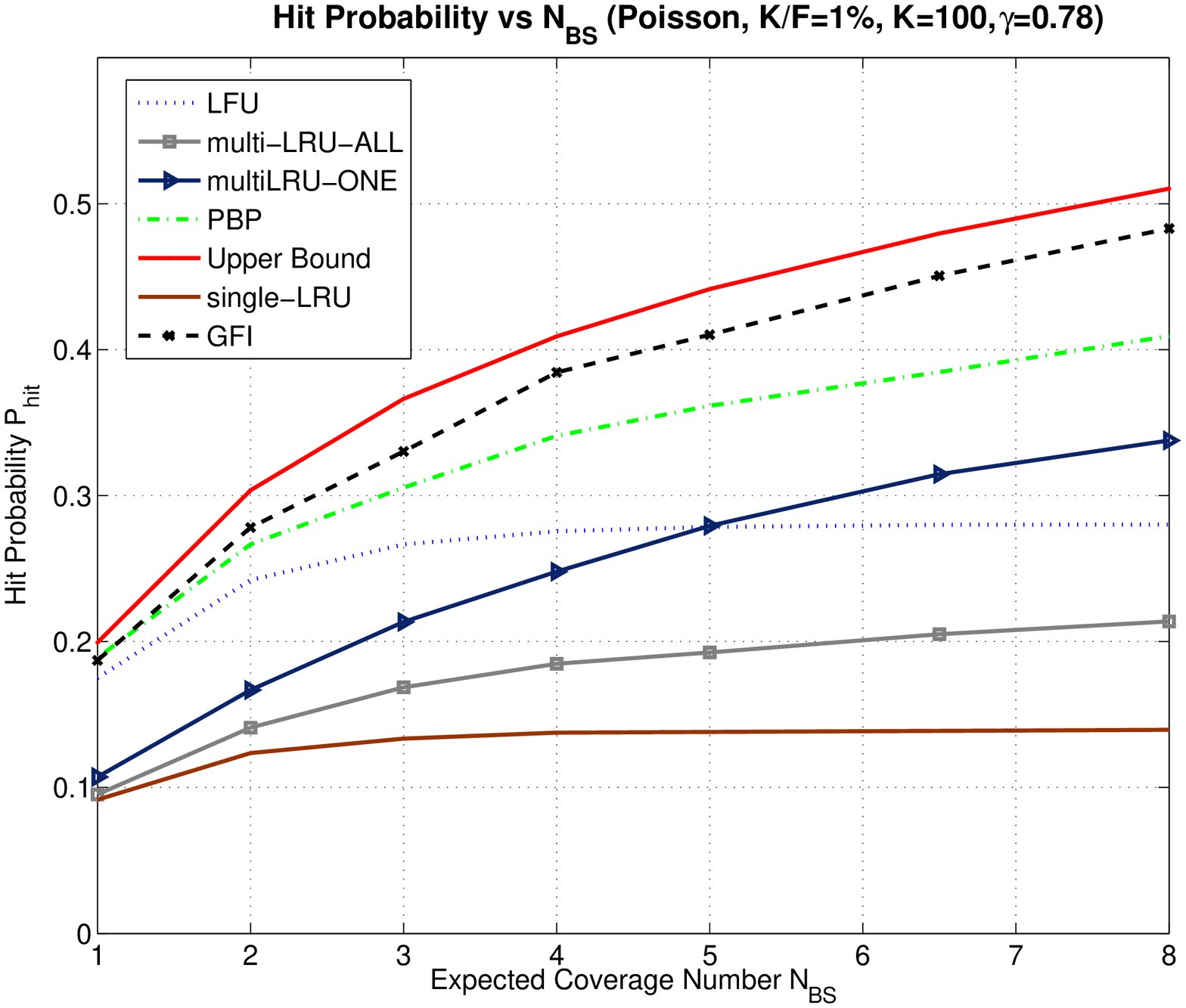,width=3.2in}
    \caption{Hit Performance PPP/Boolean, $\alpha=1\%$.}
    \label{fig:PoissonK/F=1}
%
	\centering  
    \epsfig{file=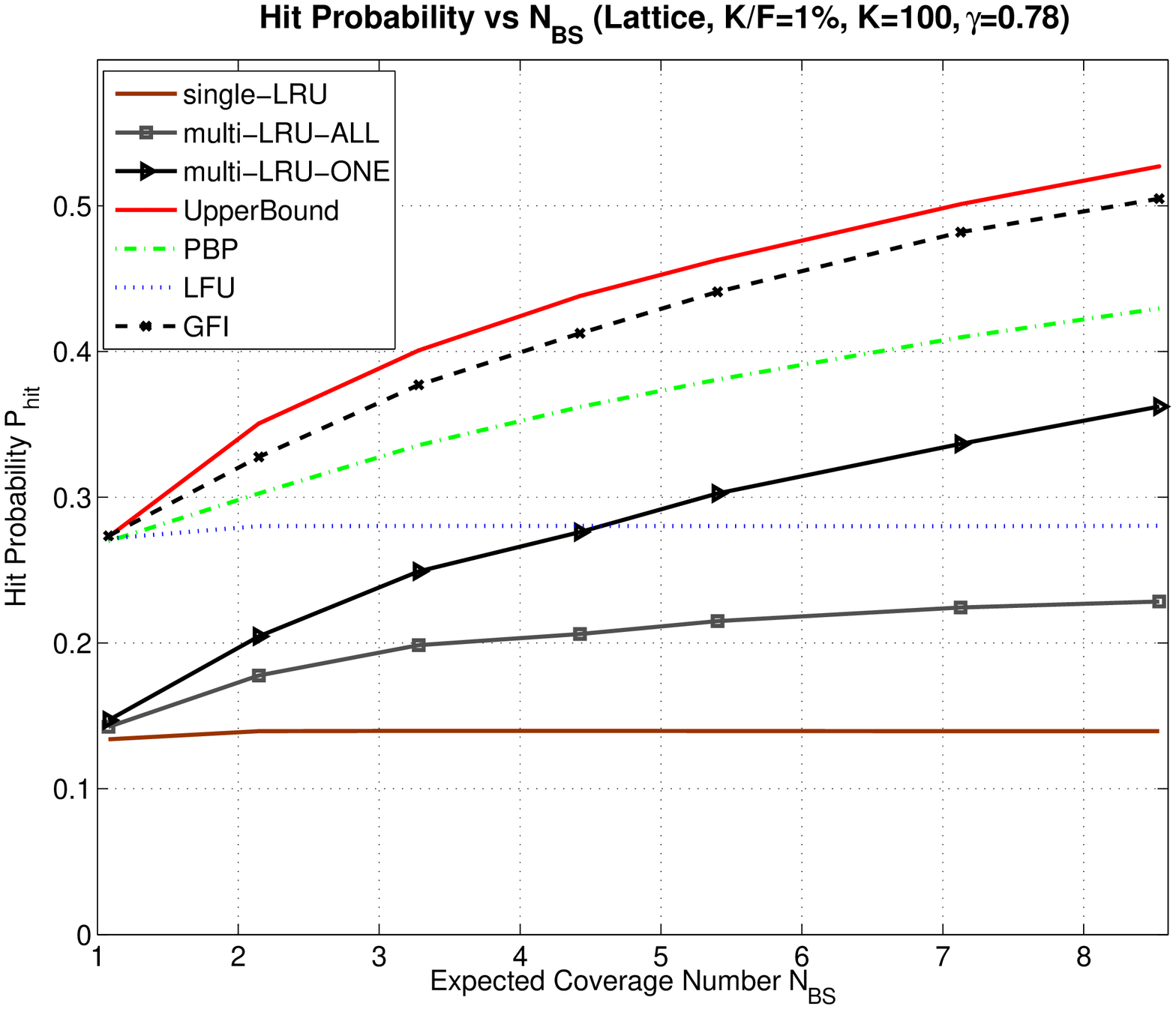,width=3.2in}
    \caption{Hit Performance Lattice/Boolean, $\alpha=1\%$.}
     \label{fig:LatticeK/F=1}
     	\centering  
    \epsfig{file=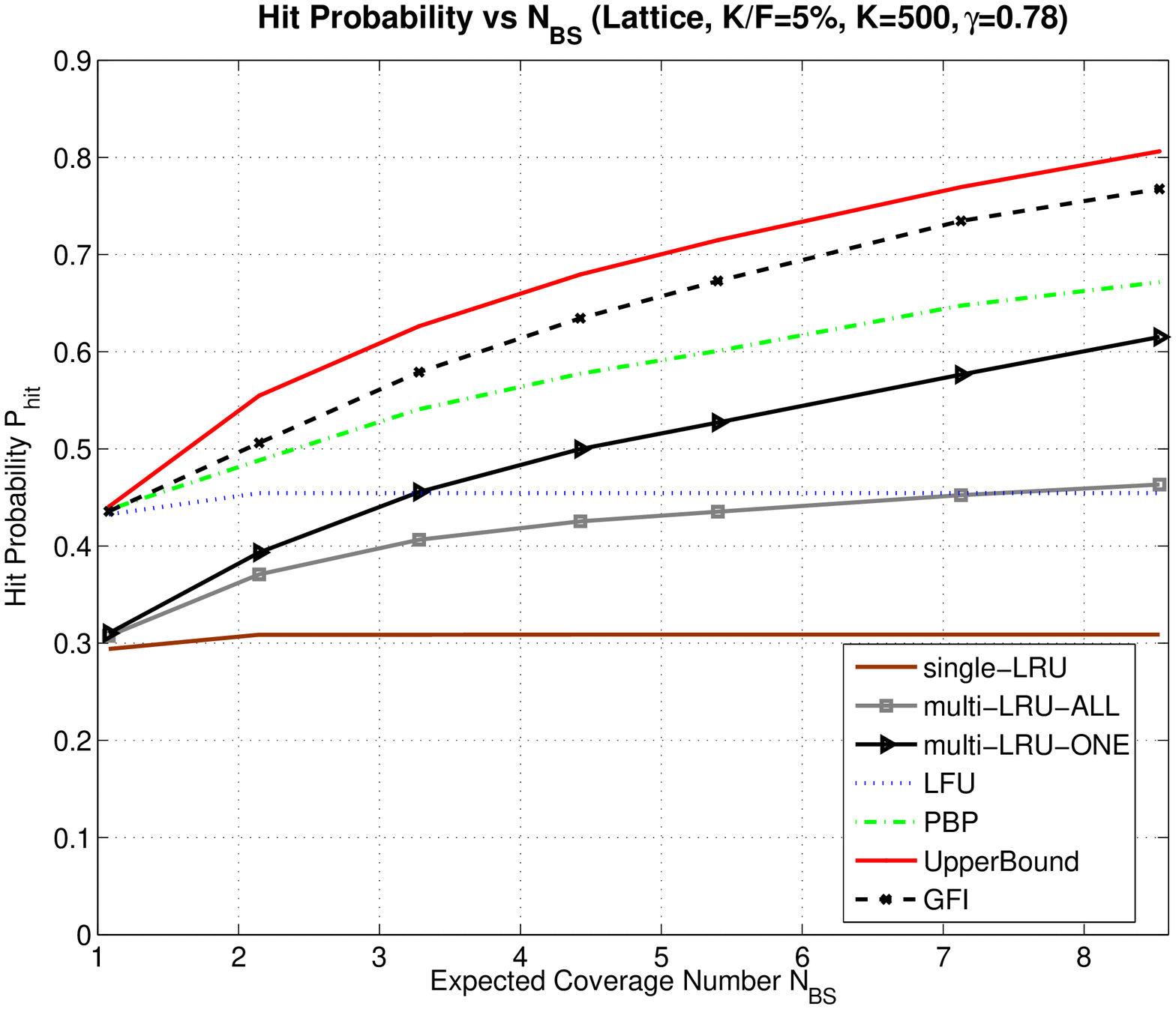,width=3.2in}
    \caption{Hit Performance Lattice/Boolean, $\alpha=5\%$.}
     \label{fig:LatticeK/F=5}
\end{figure}

In Fig. \ref{fig:PoissonK/F=1}, \ref{fig:LatticeK/F=1} and \ref{fig:LatticeK/F=5} we evaluate the hit probabilities of the proposed \textit{multi-LRU} policies over the expected number of covering stations. In the simulations the radius of the Boolean model is increased from $R_b=0.6$ to $2.25$. The radius is mapped to the expected coverage number, as in Table \ref{tab:Radius_Nexp}. In Fig. \ref{fig:PoissonK/F=1} transmission nodes are positions as a PPP, while in Fig. \ref{fig:LatticeK/F=1}, \ref{fig:LatticeK/F=5} on a Lattice. We compare the multi-LRU-One/All performance to different existing policies mentioned in this paper, like LFU, single-LRU, PBP and GFI, as well as the upper bound given in (\ref{UBhit}). The parameter $\alpha$ is chosen equal to $1\%$ in Fig.  \ref{fig:PoissonK/F=1}, \ref{fig:LatticeK/F=1} and $5\%$ in Fig. \ref{fig:LatticeK/F=5}.

As a reminder, the single-LRU policy is not influenced by multi-coverage. Each user can contact a single station, the one closest to the user. If the user request is cached in this memory, then there is a hit, otherwise  the 
object is fetched from the core network and inserted to the station's cache. 

From the three figures very interesting conclusions about the policies can be derived:

(i) Even for small values of coverage overlap (expected coverage number) a considerable increase in hit probability is achieved by using the multi-LRU policies, compared to the single-LRU. For example, when $\overline{N_{BS}}=2$, the multi-LRU-One is $42\%$ (relative gain) above the single-LRU for Lattice placement and 35\% for PPP placement. A further increase of $\overline{N_{BS}}$ makes the gain even more apparent. For  $\overline{N_{BS}}=3$ the relative gains are 70\% and 60\%, respectively.

 (ii) For every value of $\overline{N_{BS}}$ the multi-LRU-One policy performs better that the multi-LRU-All, in all three figures. This is because the same object is inserted in all memories of stations covering a user, and thus adjacent stations have similar content repositories (CSA). Consequently, a request falling in areas of overlap profits less by the diversity of content from the multiple stations that can cover her/him and the hit-performance is reduced. Although multi-LRU does not optimally leave copies in memories of neighbouring stations, it does perform much better that other policies that do not consider at all multi-coverage, like the single-LRU.
 
 (iii) From both figures, the difference between POP and POQ policies is evident. In the first group we find the POP policies \{LFU, PBP, GFI\}, while in the other one the POQ \{multi-LRU-One/All, simple LRU\}. POP policies have greater performance by exploiting the "expensive" information of known object popularity, and also the IRM assumption that this is constant over time. In a realistic environment however, where traffic patterns change over time, such policies will demand regular updates and are approximative, because they depend on the validity of the estimation over the popularity values. On the other hand, the multi-LRU policies introduced here do not depend on such information. (We currently work on the performance of the multi-LRU policies when traffic exhibits temporal locality). A related interesting remark is that, as the $\alpha$ ratio (memory-to-catalogue) increases, the difference between the two groups' performance decreases. This can be observed by comparing Fig. \ref{fig:LatticeK/F=5} to Fig. \ref{fig:LatticeK/F=1} (Lattice).
  
(iv) For $\overline{N_{BS}}$ close to 1, a user can connect to at most one station, and the performance of multi-LRU-One/All, and single-LRU coincide. The same applies for the group LFU, PBP and GFI. For $\overline{N_{BS}}\approx 1$ these last three policies tend to cache the K most popular objects in each station (LFU is doing this exactly). Hence, when a user connects to a single station then she/he gets the maximum hit probability and the upper bound also coincides.

(v) It is obvious that the two standard policies single-LRU and LFU exhibit constant performance as the multi coverage event increases, because the memory of each station is updated independently of the others and a user can be served by at most one station. A small increase of hit probability is observed for small $\overline{N_{BS}}$, because for small coverage radius, areas of no coverage exist, so that the hit probability appears reduced.

 (vi) GFI performs best among all policies, and its performance is very close to the upper bound. The latter is an indication that the upper bound is fairly tight. The good performance of the GFI comes at the cost of both a very high computational complexity for the memory allocations, as well as a considerable amount of information availability. In general, GFI is a centralised solution that requires complete knowledge over the transmission node and user positions, and over the popularity distribution. Thus, it is reasonable to surpass PBP, which uses less information (only the coverage number distribution) and less computational complexity to find the optimal memory allocations.

 (vii) For all policies, the performance is higher in the Lattice case compared to the PPP placement. The reason is due to the random placement of the PPP which can often leave areas uncovered, or covered by a number of stations smaller than the expected one. When a user demands for unpopular objects, there are less chances that these will be hit in the PPP case than in the Lattice one, due to the randomness in BS placement.
   
%


\subsubsection{q-LRU}

\begin{figure}[h!] 
	\centering  
    \epsfig{file=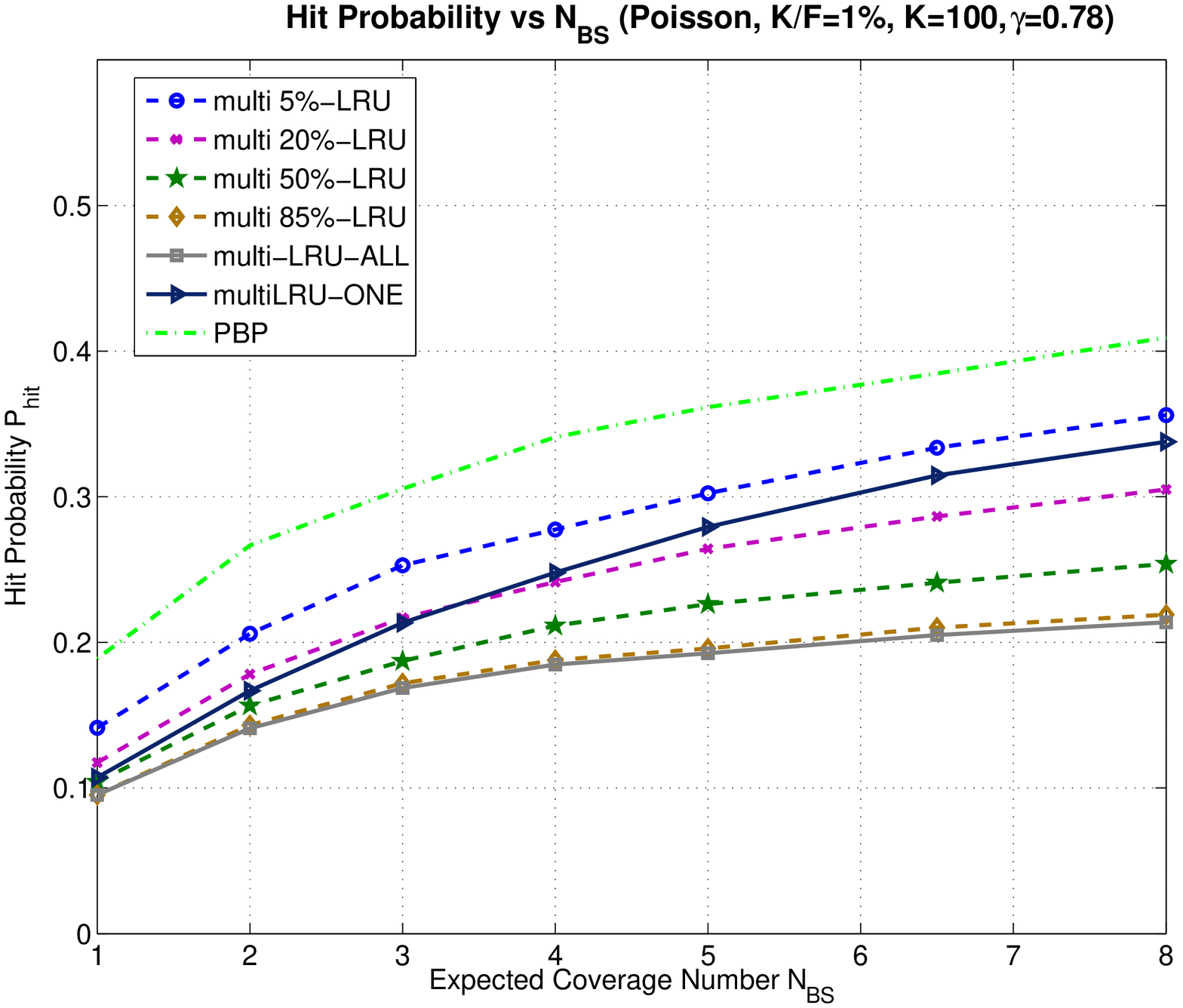,width=3.25in}
    \caption{q-multi-LRU-All, Boolean/PPP, $\alpha=1\%$.}
     \label{fig:PoissonK/F=1_qLRU}
     	\centering  
    \epsfig{file=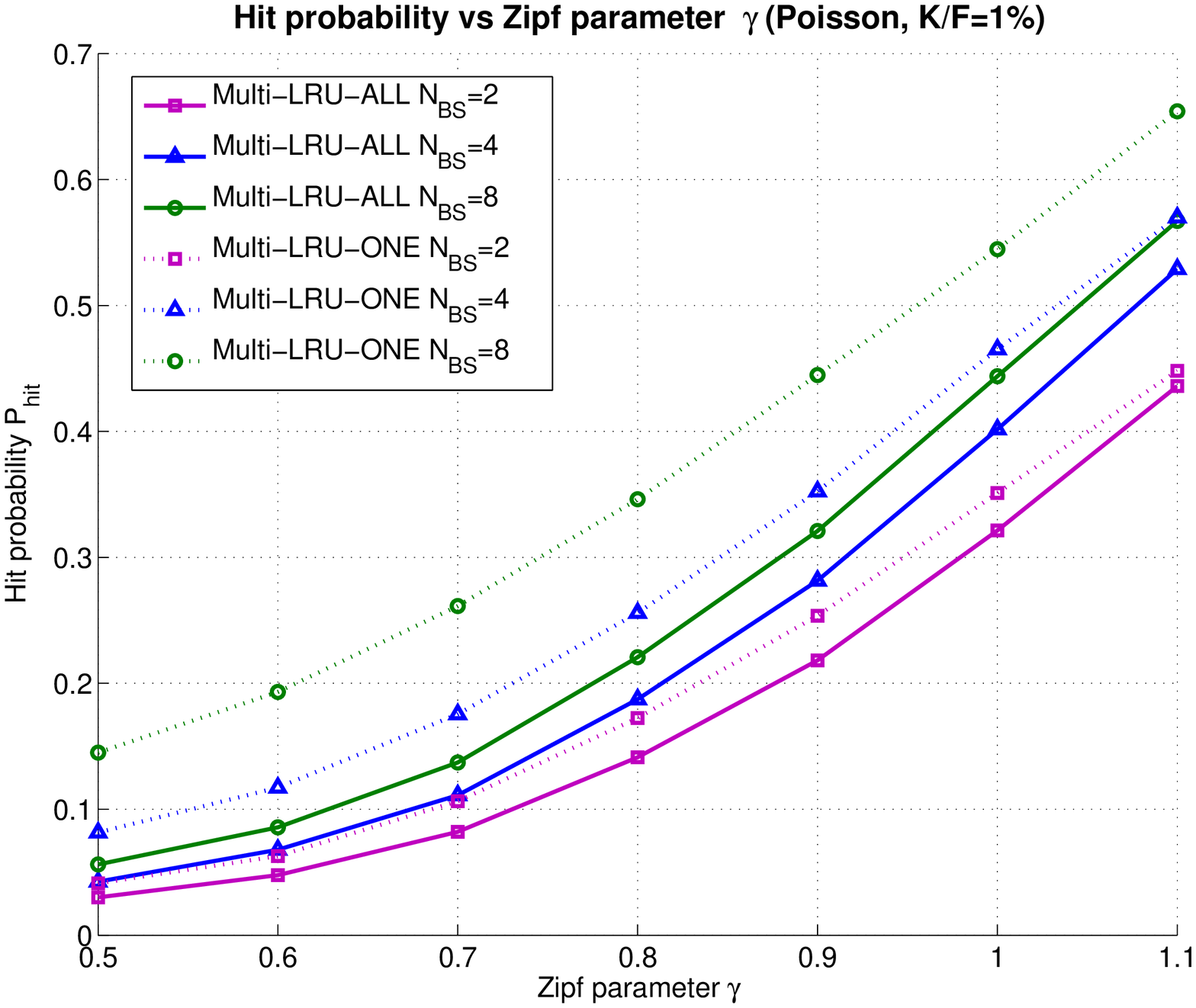,width=3.25in}
    \caption{Hit probability over $\gamma$, PPP, $\alpha=1\%$.}
     \label{fig:VaryGamma_K/F=1}
     	\centering  
    \epsfig{file=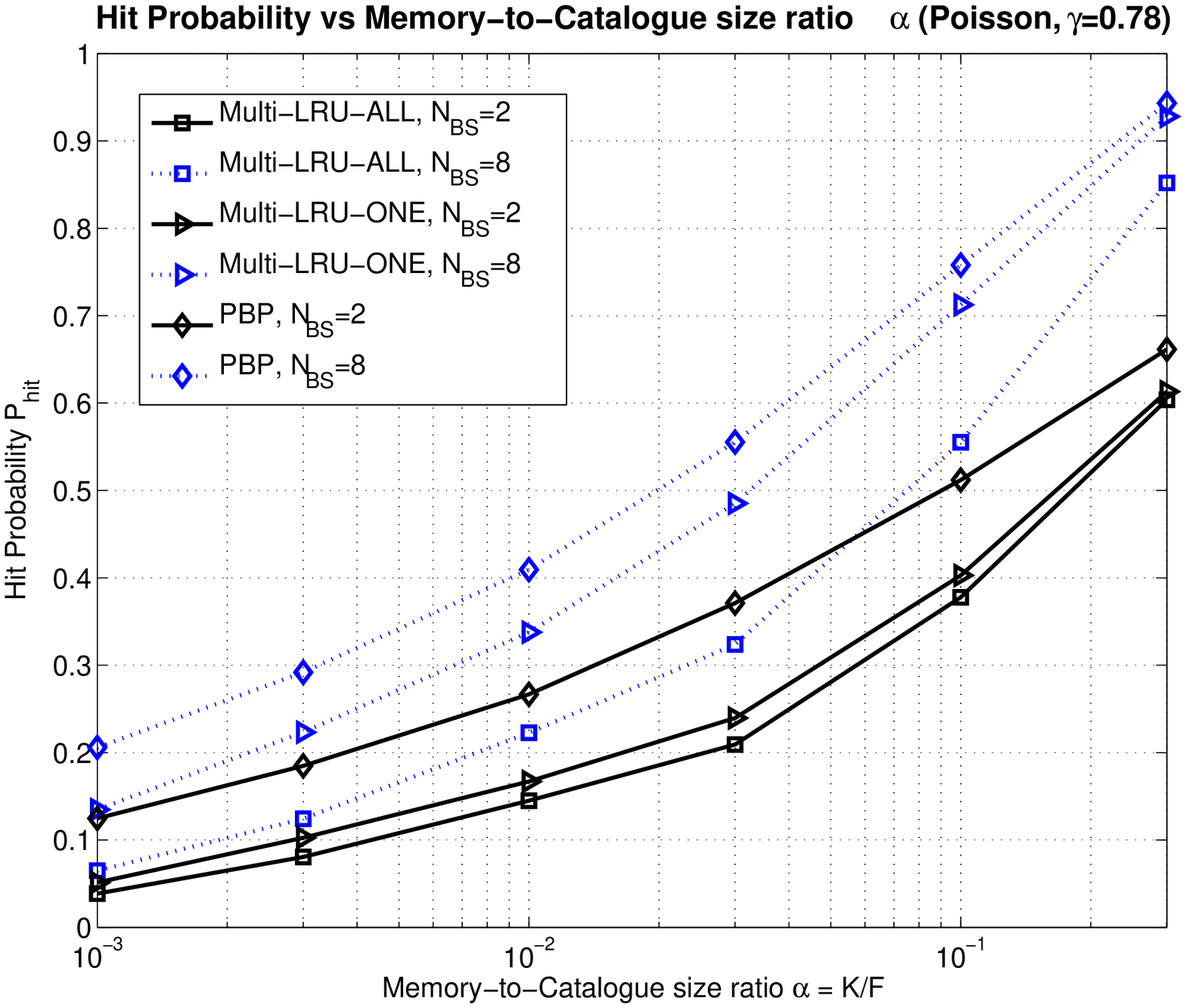,width=3.25in}
    \caption{Hit probability over $\alpha$, PPP, $\gamma=0.78$.}
     \label{fig:VaryK/F}
\end{figure}

Fig. \ref{fig:PoissonK/F=1_qLRU} plots the hit probability of q-multi-LRU-All policies for various values of $q\in\left(0,1\right]$. As in the previous figures, $\gamma=0.78$, $F=10,000$, $\lambda_b=0.5$ and stations are
modelled by PPP and have memory $K=100$.

When $q=1=100\%$, q-multi-LRU-All performs identically to the multi-LRU-All policy. As $q$ increases, the performance 
of q-multi-LRU-All monotonically decreases to that of multi-LRU-All. From a previous remark for the comparison between multi-LRU-All and multi-LRU-One, we remind the reader that caching an object in as few as possible stations prevents adjacent stations - with overlapping coverage areas - from having large similarities between their content repository. For IRM traffic, the strategy of inserting different content in neighbouring stations with common coverage areas increases the hit probability. Consequently, as
$q$ decreases, the performance of q-multi-LRU-All improves, but at the same time memories insert new content more rarely. In this sense, the good performance of q-multi-LRU-All with small $q$ exploits the IRM characteristic of stationary traffic, and will converge to good performance after a long transient period, which is often not possible for realistic traffic that exhibits faster variations in popularity and catalogue content.

\subsubsection{Zipf parameter $\gamma$}
To evaluate the impact of the Zipf parameter $\gamma$ we provide plots for the hit probability versus this parameter in Fig. \ref{fig:VaryGamma_K/F=1}. Letting $\gamma$ increase results in a popularity distribution where a small number of objects is considerably more popular than the rest of the catalogue. Eventually, hit probability will increase for both multi-LRU policies, because due to the Update phase, popular objects tend to be kept cached in memory once inserted. Furthermore, the relative difference $\frac{P_{hit}(multi-LRU-One)-P_{hit}(multi-LRU-All)}{P_{hit}(multi-LRU-All)}$ decreases as $\gamma$ increases. This happens because for increasing $\gamma$ unpopular objects have less influence on the hit probability. Consequently clever geographic placement plays less of a role to get high performance, as long as every user can find the most popular objects cached in a nearby station.
%
%

\subsubsection{Memory-to-Catalogue-size ratio $\alpha$}
Fig. \ref{fig:VaryK/F} illustrates  the behaviour of the three policies \{multi-LRU-One, multi-LRU-All, PBP\} when varying the Memory-to-Catalogue-size ratio $\alpha$ (here a larger size catalogue of $F=20,000$ is used, in order to evaluate for very small values of the $\alpha = K/F$ ratio). 
From the plots the hit probability performance is shown to increase when the ratio  $\alpha$ increases, and tends to $100\%$ as the ratio tends to 1. Furthermore, the need for smart memory allocations is less important for large values of the ratio $\alpha=K/F$ because the sum popularity of files left outside the caches is not considerable. Thus, we reasonably see in the figure, that different policies tend to have the same performance for larger values of the ratio $\alpha$.

%
%

\subsection{Traffic with temporal locality}
\label{Sec:5_4_TempLoc}

The evaluation up to this point has been restricted to IRM user (request) traffic. It can be observed in the simulation figures that the multi-LRU-One performs better than the multi-LRU-All. This is because comparison is made under the IRM, which is stationary, so, by letting the simulations run for a long time period, the performance of the multi-LRU-One can converge to high hit probabilities. This however is not true for traffic that exhibits temporal locality. Such traffic considers a finite lifespan per object, so that after a random time interval of finite expected value, the object ceases to interest users, whereas during its lifespan, its popularity may vary. Models for this type have already been proposed in \cite{TraversoTranMult15,OlmosTEMPO14}. 

\begin{figure}[ht!]   
	\centering  
	 \epsfig{file=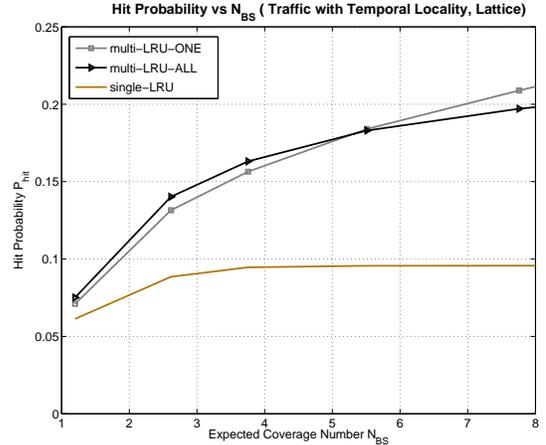,width=3.2in}
    \caption{Lattice/Boolean, Temporal locality}
     \label{fig:Temporal_multi-LRU}
\end{figure}

Based on these, we simulate a finite rectangular area with $20$ stations on a lattice. The coverage model is Boolean, and the radius is allowed to vary as before. Each station is equipped with memory of capacity $K=600$ objects. A time period $T=30$ $[months]$ is considered, where new objects appear with a rate $\lambda_{obj}=240$[\textit{objects/day}]. Each object $j$ has a lifespan $\tau_j$ from a random distribution, with average lifespan $\bar{\tau}=\mathbb{E}[\tau]=100$ \textit{days}. Requests for all objects appear with a total rate of $\lambda_u=4000$ [\textit{requests/day}] and are placed uniformly inside the simulated area. The average popularity per object is computed as $\bar{P}=\lambda_{u}/\lambda_{obj}\approx 16.7$ [\textit{requests/object}]. The evaluation of the two multi-LRU versions and the single-LRU is shown in Fig. \ref{fig:Temporal_multi-LRU}.

From the figure, it can be observed that the multi-LRU-All has a higher performance than the multi-LRU-One, until an average coverage number around six stations, which is fairly large. The reason for such change of behaviour compared to the IRM, is due to the temporal locality. An object can be hit only during its finite lifespan, so an increased number of its replicas in the network (as in the multi-LRU-All case), increases the chances to be hit before extinction. Although more replicas give less choice to a user covered by more than one station, they do bring in this case a better result. 

After a certain value of $\overline{N_{BS}}$, the multi-LRU-All performs poorer compared to -One, because of the finite memory and the fact that a single object is made available to more caches than necessary. Hence there is a limit up to which multi-LRU-All is preferable, because the memories should offer a sufficient amount of content diversity to the users. 

\section{Conclusions}

In this work a novel family of spatial multi-LRU policies is introduced, which exploit multi-coverage events of wireless networks to increase the hit probability. Two main variations are investigated, the multi-LRU-One and the -All. Che-like approximations give results close to simulation values. The multi-LRU-One provides higher object diversity in neighbouring caches and performs better under IRM traffic. The multi-LRU-All instead, lets objects quickly spread geographically and makes them immediately available to many users. This variation is profitable for traffic with temporal locality. Hence, depending on the incoming traffic either policy can be recommended. Future work should explain more clearly how the performance of these policies is affected by the spatial and temporal locality characteristics of traffic.


%
\bibliographystyle{abbrv}

%
%

\newpage
\appendix

\section{Coverage models}
\label{app:A}

1) \textit{$\mathrm{SINR}$ Model}: The quality of coverage at a planar point $y\in\mathbb{R}^2$ served by node $x_i\in\Phi_b$, is described by  the Signal-to-Interference-Noise-Ratio, $\mathrm{SINR}(y,x_i)$. Coverage at $y$ depends on node $x_i$'s power $P$ [Watt], the characteristics of the wireless channel such as fading and shadowing described by the random variable $S_i$, noise power $W$ [Watt] at the receiver $y$, as well as the total interference at $y$, written as a sum of received signals from all atoms, $I(y)= \sum_{x_i\in\Phi_b} S_i/\ell(d(y,x_i))$. Here, $d(y,x_i)$ is the Euclidean distance between receiver and transmitter. The path-loss function can take (among others) the expression $\ell(d(y,x_i))=(Bd(y,x_i))^{\beta}$, with constants $B>0$, $\beta>2$. We define,
\begin{eqnarray}
\label{SINR}
\mathrm{SINR}(y,x_i) & := & \frac{S_i/\ell(d(y,x_i))}{W/P + I - S_i/\ell(d(y,x_i))}.\nonumber
\end{eqnarray}
The coverage cell of $x_i$ is the set of all locations $y$, such that the $\mathrm{SINR}(y,x_i)$ exceeds a predefined threshold $T>0$,
\begin{eqnarray}
\label{CovCell}
\mathcal{C}_{i}^{\mathrm{SINR}}(T) & = & \left\{y\in\mathbb{R}^2: \mathrm{SINR}(y,x_i)>T\right\}.\nonumber
\end{eqnarray} 
(For the typical location at $(0,0)$, we omit $o$ and write $\mathrm{SINR}(x_i) := \mathrm{SINR}(o,x_i)$. Furthermore, we can substitute the radial distance of $x_i$ from the origin, $d(o,x_i)=r_i$.)

The coverage number indicates how many distinct cells cover the typical location simultaneously, and is the r.v.
\begin{eqnarray}
\label{SINRcovnum}
\mathcal{N}^{\mathrm{SINR}}(T) & = & \sum_{x_i\in\Phi_b}\mathbbm{1}[\mathrm{SINR}(x_i)>T].\nonumber
\end{eqnarray}
For general shadowing conditions, the authors in \cite{KeelerBartek13} have calculated exactly the probabilities
\begin{eqnarray}
\label{pSINR}
p^{\mathrm{SINR}}_m(T) := \mathbb{P}\left[\mathcal{N}^{\mathrm{SINR}}(T)=m\right], & & \forall m.\nonumber
\end{eqnarray}

2) \textit{$\mathrm{SNR}$ and Boolean Model}: In the noise-limited case, interference is omitted, because its power is considered unimportant compared to noise. The expression in (\ref{SINR}) then simplifies to the $\mathrm{SNR}$ case,
\begin{eqnarray}
\label{SNR}
\mathrm{SNR}(y,x_i) & := & \frac{S_i/\ell(d(y,x_i))}{W/P}.\nonumber
\end{eqnarray}
Without channel variations $S_i=1$, the $\mathrm{SNR}$ coverage cell reduces to a disc of center $x_i$ and radius $R_b(T):=T^{-1/\beta}\tilde{B}^{-1}$, where $\tilde{B}=W^{1/{\beta}}B$. This is formally written as
\begin{eqnarray}
\label{CovCell}
\mathcal{C}_{i}^{B}(T) & = & \left\{y\in\mathbb{R}^2: d(y,x_i)<R_b(T)\right\}.\nonumber
\end{eqnarray} 
The coverage number is given, similarly to (\ref{SINRcovnum}), by 
\begin{eqnarray}
\label{Boolcovnum}
\mathcal{N}^{B}(T) & = & 
\sum_{i\in\mathbb{N}_+}\mathbbm{1}[r_i<R_b(T)].\nonumber
\end{eqnarray}

From \cite[Lemma 3.1]{BacBlaVol1} (or \cite[Th. 13.5]{HaenggiBook13}) we know for the Boolean model  that $\mathcal{N}^B(T)$ is Poisson distributed with parameter $\nu(T):=\lambda_b\pi R_b(T)^2=\lambda_b\pi T^{-2/\beta}W^{-2/\beta}B^{-2}$, so,
\begin{eqnarray}
\label{pBool}
p_m^B(T) = \frac{\nu(T)^m}{m!}e^{-\nu(T)}, & & \forall m.
\end{eqnarray}

%
\begin{figure*}[t!]   
\centering  
\subfigure[multi-LRU-One]{
           \epsfig{file=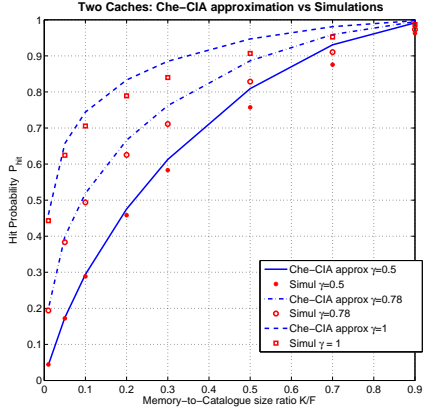, width=2.5in}
           \label{Che2:One}
           }
           \subfigure[multi-LRU-All]{
	   \epsfig{file=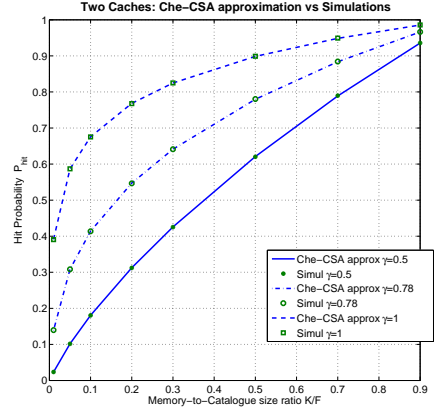, width=2.5in}
	   \label{Che2:All}   
           }
         \caption{Che approximation for (a) multi-LRU-One and (b) multi-LRU-All in the two-cache network. Hit probability versus memory-to-catalogue size ratio $a=K/F$, $F=10,000$ objects, for different Zipf parameter $\gamma$.}
         \label{Che2cache}
\end{figure*} 

\section{multi-LRU: two-cache network}
\label{app:B}

To understand how the Che-like approximations work for the multi-LRU policies, we analyse the simple network of two nodes $x_i$, $i\in\left\{1,2\right\}$, each one equipped with a cache of size $K$. 
Each node covers an entire area $A\subset\mathbb{R}^2$, so that all planar points are covered by both nodes. The total area is divided in two Voronoi cells $\mathcal{V}(x_i)$. To simplify further, we assume equal-sized Voronoi cells $|\mathcal{V}(x_1)|=|\mathcal{V}(x_1)|=|\mathcal{V}|$. 


We apply the analysis of Section \ref{GenChe} to this network model. Specifically, the formula for the hit probability of an object $c_j$ at cache $\Xi_i$ in (\ref{Phitgen2}), takes the expression (for $i=\{1,2\}$),
\begin{eqnarray}
\label{Phiti2}
P_{hit,i}\left(j\right) & \stackrel{IRM}{=} & \mathbb{P}\left(u_{-1}\in\left(\mathcal{S}_{-1},\left| t_{o}-t_{-1}\right|<T_{C},j\right)\right)\cdot \nonumber\\
& & \cdot \left[P_{hit,i}\left(j\right)+\mathbb{P}\left(c_j\notin\Xi_1\cap c_j\notin\Xi_2\right)\right].
\end{eqnarray} 
Solving the above over $P_{hit,i}\left(j\right)$ gives an expression for the hit probability of object $c_j$ at cache $\Xi_i$. The characteristic time $T_{C}$ is found by solving the equation (\ref{TCgen}),
\begin{eqnarray}
\label{Kincache}
\sum_{j=1}^F P_{hit,i}\left(j\right)=K, & i=\{1,2\}.
\end{eqnarray}
Finally, the total hit probability (\ref{PhitTOTgen}) takes both caches into account, and is equal to 
\begin{eqnarray}
\label{HitTot2}
P_{hit} & = & \sum_{j=1}^F a_j \left(1-\mathbb{P}\left(c_j\notin\Xi_1\cap c_j\notin\Xi_2\right)\right).
\end{eqnarray}

$\bullet$ \textbf{multi-LRU-One}: In this case, $\mathcal{S}_{o}=\mathcal{S}_{-1}=\mathcal{V}(x_i)$, in (\ref{Phiti2}). Table \ref{Tab2a} gives all pairs of inventory states that a user $u_o$ arriving at $t_o^{-}$ sees, when the previous user $u_{-1}$ asking for the same content arrived in cell (say) $\psi_{-1}\in\mathcal{V}(x_1)$ at some time $t_{-1}^{-}$, such that $\left|t_o-t_{-1}\right|\leq T_C$. We denote by logical $1$ the fact that the object is in the cache and by $0$ otherwise. From the table it is clear that user $u_{-1}$ does not take any action on cache $\Xi_2$, this is why, when $\mathbbm{1}\left[c_j\in\Xi_2(t_{-1}^{-})\right]=1$, we cannot know whether the content will remain in the cache till $t_o^{-}$, so we write $\mathbbm{1}\left[c_j\in\Xi_2(t_{o}^{-})\right]\in\left\{0,1\right\}$.
\begin{table}[ht!]
\caption{multi-LRU-One: States at $t_{-1}^-$ and $t_{o}^-$}
\centering
\begin{tabular}{| c | c | c | c | c | c | }
\hline
$\Xi_1(t_{-1}^{-})$ & $\Xi_2(t_{-1}^{-})$ &  & $\Xi_1(t_{o}^{-})$ & $\Xi_2(t_{o}^{-})$ & \\
\hline
0 & 0 & $\rightarrow$ & 1 & 0 & insert 1\\
0 & 1 & $\rightarrow$ & 0 & $\left\{0,1\right\}$ & no update\\
1 & 0 & $\rightarrow$ & 1 & 0 & update 1\\
1 & 1 & $\rightarrow$ & 1 & $\left\{0,1\right\}$ & update 1\\
\hline
\end{tabular}
\label{Tab2a}
\end{table}

There is the unknown probability $\mathbb{P}\left(c_j\neq \Xi_1\cap c_j\neq \Xi_2\right) = 1- \mathbb{P}\left(c_j \in\Xi_1\cup c_j\in\Xi_2\right)$. For multi-LRU-One, we observe that an insertion of an object is triggered when its request arrives but does not find the object inside any of the two caches. However, the insertion is done only in the closest cache and stays there for time $T_C$. During this time, the same object cannot be inserted in the other cache, hence, $\left\{c_j\in\Xi_1\right\}$ and $\left\{c_j\in\Xi_2\right\}$ are mutually exclusive events. Then,
\begin{eqnarray}
\label{jneq12one2}
\mathbb{P}\left(c_j\notin\Xi_1,c_j\notin\Xi_2\right) & = & 1 - \mathbb{P}\left(c_j \in\Xi_1\cup c_j\in\Xi_2\right)\nonumber\\ 
& = & 
1 - 2P_{hit,1}(j),
\end{eqnarray}
where the last equality is due to the symmetry of our model and the IRM traffic. However, in more general cases of node placement and coverage, content exclusivity is not true, because only a small area of the coverage cell will overlap with one neighbour. Users in other areas of the cell will be covered by other neighbours that can trigger the insertion of the same object, anyway. Hence, this result is not of much use for the PP coverage models. For this reason we want to evaluate how the CIA approximation applies here. For the two-cache model, this means for $\Xi_1$ (or $\Xi_2$),
\begin{eqnarray}
\label{CIA22}
\mathbb{P}\left(c_j\notin\Xi_1\cap c_j\notin\Xi_2\right) 
& = & 
 1 - P_{hit,1}(j), \ \ (CIA_1).
\end{eqnarray}
We can then replace in (\ref{Phiti2}) and (\ref{Kincache}) to get (for $i\in\left\{1,2\right\}$)
\begin{eqnarray}
\label{PhitOne2Ia}
P_{hit,i}(j) & = &  1-e^{-a_j\lambda_u|\mathcal{V}|T_C},\\
\label{PhitOne2Ib}
\sum_{j=1}^F P_{hit,i}(j) & = & \sum_{j=1}^F \left(1-e^{-a_j\lambda_u|\mathcal{V}|T_C}\right) = K.
\end{eqnarray}
For the total $P_{hit}$ probability, we should appropriately adapt the form in (\ref{HitTot2}) to the $CIA_2$ approximation,
\begin{eqnarray}
\label{PhitCIA2}
P_{hit} & = & \sum_{j=1}^F a_j \left(1-(\mathbb{P}\left(c_j\notin\Xi_1\right))^2\right)\nonumber\\
& = & \sum_{j=1}^F a_j (1-e^{-a_j\lambda_u2|\mathcal{V}|T_C}),
\end{eqnarray}
and the area $2|\mathcal{V}|=|A|$ is equal to the total coverage cell.


$\bullet$ \textbf{multi-LRU-All}: In this case, $\mathcal{S}_{-1}=\mathcal{S}_o= A$ in (\ref{Phiti2}) , for the hit probability of node $i$. 

To calculate the unknown probability $\mathbb{P}\left(c_j\notin\Xi_1\cap c_j\notin\Xi_2\right)$ we argue as follows. 
In the case of multi-LRU-All, an object cannot be inserted in cache 1 if not inserted also in cache 2 and the other way round. Based on the Che approximation, once the object is inserted it stays $T_C$ amount of time, before removed from each cache. Hence, the existence of an object in one cache implies the existence of the same object in the other. So, due to the model's symmetry
\begin{eqnarray}
\label{jneq12all2}
\mathbb{P}\left(c_j\notin\Xi_1\cap c_j\notin\Xi_2\right) & = & 1 - \mathbb{P}\left(c_j \in\Xi_1\cup c_j\in\Xi_2\right)\nonumber\\ 
& = & 
1 - P_{hit,1}(j).
\end{eqnarray}
This is simply the \textit{Cache Similarity Approximation (CSA)}, which for the two-cache network is exact! Then (\ref{Phiti2}) gives,
\begin{eqnarray}
\label{PhitiAll2}
P_{hit,i}(j) & = &
1-e^{-a_j\lambda_u|A|T_C}.
\end{eqnarray}
To find the characteristic time, we need to solve (\ref{Kincache}), 
\begin{eqnarray}
\label{TCAll2}
\sum_{j=1}^F P_{hit,i}(j)= \sum_{j=1}^F \left(1-e^{-a_j\lambda_u|A|T_C}\right) = K.
\end{eqnarray}
The total hit probability is equal to,
\begin{eqnarray}
\label{HitTotAll2}
P_{hit} & = & \sum_{j=1}^F a_j \left(1-\mathbb{P}\left(c_j\notin\Xi_1,c_j\notin\Xi_2\right)\right)\nonumber\\
& \stackrel{(\ref{PhitiAll2})}{=} & \sum_{j=1}^F a_j (1-e^{-a_j\lambda_u|A|T_C}).
\end{eqnarray}

An \textit{alternative way} to calculate $\mathbb{P}\left(c_j\notin\Xi_1(t_{o}),c_j\notin\Xi_2(t_{o})\right)$ is the following. A user $u_{o}$ finds the two caches without object $c_j$, if the previous user $u_{-1}$ (at say $\mathcal{S}_{-1}=\mathcal{V}(x_1)$) with the same demand, arrived either (i) at $t_{-1}^-:\left|t_{o}-t_{-1} \right|>T_C$, so that whatever the state of the two caches $\Xi_1(t_{-1}^-)$, $\Xi_2(t_{-1}^-)$, the object $c_j$ is eventually removed, since more than $T_C$ elapsed till $t_{o}$, or (ii) at $\left|t_{o}-t_{-1} \right|\leq T_C$. In the second case all possible change of states for the two caches is shown in Table \ref{Tab2b}. From this, we note that, the object will always be found in at least one of the two caches at $t_{o}$, so that the time difference can not be smaller than $T_C$. Hence,
\begin{eqnarray}
\label{jneq12all2alt}
\mathbb{P}\left(c_j\notin\Xi_1(t_{o}),c_j\notin\Xi_2(t_{o})\right) 
\stackrel{IRM}{=} e^{-a_j\lambda_u|A|T_C}.
\end{eqnarray}
The expressions in (\ref{jneq12all2alt}) and (\ref{PhitiAll2}) are the same.
\begin{table}[ht!]
\caption{multi-LRU-All: States at $t_{-1}^-$ and $t_{o}^-$}
\centering
\begin{tabular}{| c | c | c | c | c | c | }
\hline
$\Xi_1(t_{-1}^-)$ & $\Xi_2(t_{-1}^-)$ &  & $\Xi_1(t_{o}^-)$ & $\Xi_2(t_{o}^-)$ & \\
\hline
0 & 0 & $\rightarrow$ & 1 & 1 & insert both\\
0 & 1 & $\rightarrow$ & 0 & 1 & update 2\\
1 & 0 & $\rightarrow$ & 1 & 0 & update 1\\
1 & 1 & $\rightarrow$ & 1 & 1 & update both\\
\hline
\end{tabular}
\label{Tab2b}
\end{table}

The accuracy of the approximations in the two-cache network is shown in Fig.\ref{Che2cache}. The Che-CIA approximation for multi-LRU-One - although not accurate - performs reasonably well in the two-cache network. The Che-CSA approximation for the multi-LRU-All, is exact.

\end{document}